\documentclass[12pt]{article}
\usepackage{amssymb}
\usepackage{graphicx}
\usepackage{epsfig}
\usepackage{amsmath}
\usepackage{color}

\def\CN{{\cal N}}
\newcommand{\nn}{\nonumber}

\newtheorem{theorem}{Theorem}

\newtheorem{corollary}[theorem]{Corollary}

\newtheorem{definition}[theorem]{Defintion}

\newtheorem{lemma}[theorem]{Lemma}

\newtheorem{remark}[theorem]{Remark}

\newenvironment{proof}[1][Proof]{\textbf{#1.} }{\ \rule{0.5em}{0.5em}}
\newcommand{\Z}{\mathbb{Z}}
\newcommand{\C}{\mathbb{C}}
\newcommand{\R}{\mathbb{R}}
\newcommand{\HH}{\mathbb{H}}

\newcommand{\g}{\frak{g}}
\newcommand{\h}{\frak{h}}

\newcommand{\msp}{{\mathcal M}_{\rm coassoc}}

\begin{document}

\begin{titlepage}

\begin{center}
\hfill HUTP-02/A005\\
\vskip 0.8 cm
{\LARGE Duality and Fibrations on $G_2$ Manifolds}

\vskip 0.8 cm
{\small
\textsc{Sergei Gukov,$^{*}$ Shing-Tung Yau,$^{**}$
and Eric Zaslow$^{***}$}
\vskip 0.5cm
$^{*}$\textsc{\sl Department of Physics, Harvard University}
\vskip 0.1cm
$^{**}$\textsc{\sl
Department of Mathematics, Harvard University}
\vskip 0.1cm
$^{***}$\textsc{\sl Department of Mathematics, Northwestern University}}

\end{center}

\vskip 0.5 cm
{\small
\begin{abstract}
We\footnotetext{\tt gukov@democritus.harvard.edu,
yau@math.harvard.edu, zaslow@math.northwestern.edu}
argue that $G_2$ manifolds for M-theory admitting
string theory Calabi-Yau duals are
fibered by coassociative submanifolds.
Dual theories are constructed using the
moduli space of M5-brane fibers as target space.
Mirror symmetry and various string and M-theory
dualities involving
$G_2$ manifolds
may be incorporated into this framework.
To give some examples, we construct two non-compact manifolds
with $G_2$ structures:  one with a $K3$ fibration,
and one with a torus fibration and a metric of $G_2$ holonomy.
Kaluza-Klein reduction of the latter solution
gives abelian BPS monopoles in $3+1$ dimensions.
\vskip0.5cm
{\scriptsize{\tableofcontents}}
\end{abstract}
}

\end{titlepage}

\pagestyle{plain}
\setcounter{page}{1}
\newcounter{bean}
\baselineskip16pt

\section{Introduction}

One of the main achievments in
string theory during the last decade
was the discovery of string dualities and relations among them.
A particularly rich and interesting
example of string duality
is {\it mirror symmetry} between pairs of
Calabi-Yau manifolds.
A geometric framework for understanding
this duality was proposed in \cite{SYZ},
and involves constructing the mirror manifold
by dualizing a torus fibration.
This construction arose from
the correspondence among nonperturbative states
of dual theories.
M-theory has united the disparate string theories
and promises to reveal the nature of string dualities.
In M-theory, the analogue
of a Calabi-Yau manifold is a manifold
with $G_2$ holonomy,
simply by the counting of dimensions:
what was $10=4+6$ for string theory is $11=4+7$ in M-theory.
According to this simple formula,
seven-dimensional $G_2$-holonomy manifolds
are natural candidates for minimally supersymmetric
(and phenomenologically interesting \cite{AchW})
compactifications of M-theory to
$3+1$ dimensions.
If manifolds with $G_2$ holonomy
are M-theory analogues of Calabi-Yau
spaces, then what is the corresponding
notion of mirror symmetry, and what
is the geometry behind duality?
Is there a fibration structure on $G_2$ manifolds relevant
to
this and possibly other string/M-theory dualities?
These are the questions that one might naturally ask, and that
we attempt to address in this paper.

We argue that, just as Calabi-Yau manifolds involved in mirror
symmetry are fibered by special Lagrangian tori, in M-theory
$G_2$-holonomy manifolds which admit string theory duals
are fibered by coassociative 4-manifolds.
Specifically, M-Theory on a seven-manifold $X,$
with $G_2$ holonomy, leads to an effective
field theory in four dimensions with $\CN=1$
supersymmetry.  The same is true for the heterotic
string theory on a Calabi-Yau manifold, $Y,$ with an
appropriate choice
of holomorphic bundle.\footnote{One requires $E\rightarrow Y$
to obey $p_1(E) = p_1(TY)$ and $c_1(E) = 0$
so that there is no anomaly, {\it i.e.} the heterotic
theory contains no fivebranes.}
Similarly, Type-IIA string theory on a
noncompact Calabi-Yau manifold,
$Z,$ with Ramond-Ramond fluxes turned on
(or with a spacetime-filling brane)
has $\CN=1$ supersymmetry
if the fluxes satisfy certain first-order equations.
Are there pairs\footnote{Kachru and Vafa first found
heterotic-Type-IIA $(Y,Z)$ pairs in \cite{KV};
some $(X,Z)$ pairs are studied in \cite{AMV,AtW}.}
$(X,Y)$ or $(X,Z)$ which lead to equivalent theories?
If so, how are the geometry and topology of $X$ related to the
choice of $Y$ or $Z,$ as well as to the
bundle or Ramond-Ramond field data?
Is there a constructive way of producing duals?

In this paper we address these questions from two
points of view, then produce two manifolds which may
serve as sources of further study in these directions:
one a torus fibration with a $G_2$ metric
constructed from Hitchin's method \cite{Hitchin} in Sec. \ref{torusfib},
and one a K3 fibration with a $G_2$ structure
(but neither closed nor co-closed three-form)
constructed in Sec. \ref{k3fib}.
The two lines of reasoning are as follows.
First, we study the moduli space of an M-theory
five-brane wrapped around a
coassociative (internal) four-cycle, $C.$
By allowing the moduli of $C$ to vary slowly in
spacetime directions, one sees
that the resulting
theory on the spacetime soliton string
is a conventional string theory with
target space
the M-brane moduli space.  The moduli space is
$Y$ or $Z,$ depending on $C.$
This line of reasoning follows \cite{HS}.
Second, as in \cite{AchW},
we use fiberwise duality
of M-theory on $K3$ with heterotic strings
on $T^3,$ as well as the ``fact'' that
Calabi-Yau's admit torus fibrations, to
argue that heterotic string theory on a Calabi-Yau
manifold should be dual to M-theory on a $K3$ fibration.

\begin{remark}
\label{otherconjs}
Some duality conjectures involving $G_2$ manifolds
have been proposed in \cite{LeeLeung} and \cite{BA}.
Our arguments don't involve pairs of $G_2$ manifolds
{\sl per se}, though do lead to relations.  For example,
if one takes a Calabi-Yau resulting from the moduli
space of a coassociative fiber, one can look for a different
$G_2$ manifold which has that Calabi-Yau as its
Kaluza-Klein reduction.  One would then expect
that two $G_2$ manifolds related in this way would
be mirror, in the sense of Shatashvili-Vafa \cite{SV}.
The setting for Acharya's arguments in \cite{BA}
is string theory
and duality of $G_2$ manifolds via dual torus fibrations.
The motivation in \cite{LeeLeung} is more mathematical,
where a fiberwise Fourier-Mukai transform leads to
conjectured dual $G_2$ manifolds near limiting
points in the moduli space of $G_2$-holonomy metrics.
\end{remark}


\begin{remark}
Results in Sec. \ref{fibbranmod} rely on physical arguments,
and include some speculative mathematics.
While the metrics of Secs.  \ref{k3fib} and \ref{torusfib}
are motivated by the physical reasoning, these sections
are purely mathematical in nature, and can be read independently.
Sec. \ref{monopole} is a mixed bag.
\end{remark}

\section{Fibrations from Brane Moduli Spaces}
\label{fibbranmod}

\subsection{Fibrations from M-theory}
\label{mfib}

The arguments of mirror symmetry as T-duality \cite{SYZ}
arise from recognizing that string duality demands
a correspondence among the nonperturbative states
of the theory.  The dual theory is then found
as a sigma model on the moduli space of a relevant brane.
We shall try to apply similar reasoning to M-theory on a compact
$G_2$ manifold, $X.$  Here, instead of D-branes
we have M-theory five-branes.\footnote{We
recall that the field content of M-theory contains
a three-form $H$ with four-form field strength; it
obeys $dH = \delta_D,$ where $D$ is the five-brane
world-volume.  This leads to a condition that the
normal bundle have trivial Euler characteristic,
which is true for the examples ($T^4,$ $K3$) in this paper.}
On a $G_2$ manifold we can choose a five-brane
whose world-volume is $\Sigma \times C,$ with
$\Sigma \subset \R^4$ a Riemann surface
in flat space, and $C \subset X$
a coassociative four-cycle
(this means $\Phi\vert_C = 0,$
where $\Phi$ is the associative calibration
three-form of the torsion-free $G_2$ structure on $X$).
The five-brane is a string in the effective theory,
the so-called {\it black string},
and we have chosen a supersymmetric brane,
in the sense that the theory on the string worldsheet is
a {\em two-dimensional} $\CN=(2,0)$ supersymmetric theory.
Its moduli space equals the moduli space of the five-brane.
Recall that similar reasoning led to the discovery
of the heterotic string as a type-II soliton \cite{HS}.

The question now arises:  What is the five-brane
moduli space? This is comprised of a choice
of coassociative 4-cycle $C \subset X,$ a choice of
$\Sigma\subset \R^4,$ and a point in the intermediate Jacobian \cite{W}:
\begin{equation}
J_{C\times \Sigma}
\equiv H^3 (C \times \Sigma , \R ) / H^3 ( C \times \Sigma , \Z )
\end{equation}
In the world-volume action, only
the self-dual three-form couples to the five-brane.
We can write a self-dual three-form as
a linear combination of
$\alpha_+ \wedge \beta_+$
and $\alpha_- \wedge \beta_-,$
where $\alpha_{\pm}$ are self-dual/anti-self-dual
forms on $\Sigma,$ and $\beta_{\pm}$ are
self-dual/anti-self-dual forms on $C.$
There are $b_{2}^{\pm}(C)$ self-dual/anti-self-dual forms on $C$.
Let us now turn to the moduli fields of
the five-brane $\Sigma\times C.$
There are two transverse directions for
$\Sigma \subset \R^4$
(or four total, as ${\rm dim} G_{\R}(2,4)=4$).
As for the number of deformations of a coassociative
submanifold \cite{M}, this is equal to $b_2^+(C)$,
and these moduli fields have left- and right- dependencies.

Now let's look at the effective theory on the black
string worldsheet, $\Sigma.$  The low-lying excitations
are described by allowing the coassociative cycle $C$
and gauge two-form on it
to vary slowly over $\Sigma.$  The effective string
theory is therefore a supersymmetric sigma model with
target space described by the $2b_2^+(C)+2$
left-moving and $b_2^+(C)+b_2^-(C)+2$ right-moving
moduli fields found above.\footnote{
The spectrum of fermions on the black string worldsheet $\Sigma$
follows by supersymmetry. Since the five-brane breaks
the six-dimensional Lorentz invariance down to:
${\rm Spin} (1,1)_{\Sigma} \times {\rm Spin} (4)_C$
the fermions in the five-brane tensor multiplet transform
as $(+, {\bf 2}_+) \oplus (-, {\bf 2}_-)$.
On the other hand, the structure group of the five-brane
normal bundle, $N$, becomes the R-symmetry group
${\rm Spin} (4)_N \cong SU(2) \times SU(2)$ of
the world-volume theory. The fermions transform in
$({\bf 1}, {\bf 2}) \oplus ({\bf 1} , {\bf 2})$
under this group \cite{BSV}.
Therefore, on the coassociative 4-manifold $C$
we have a topologically twisted $\CN=4$ gauge theory,
the so called Vafa-Witten theory \cite{VW}.
The partition function of this theory counts the euler
number of the moduli space of instantons on $C$.
It would be interesting to investigate a further relation
between this topological theory and M-theory on
$G_2$-holonomy manifolds.  Mirror symmetry
and the counting associative 3-cylces should play an important
role in such a relation.}
These fields live in a compactification defined by the
moduli space and the integer lattice $H^2(C,\Z).$
When $C \cong T^4$ we have $b_2^+ = b_2^- = 3,$ and
we find eight left-movers and eight right-movers, equal in number
to the transverse oscillations of the Type-II string.
When $C \cong K3$ we have $b_2^+=3, b_2^- = 19,$
so we get eight left-movers and 24 right-movers,
as in the heterotic string theory.
We expect these string theories to be
dual to the original M-theory on $X.$
As with SYZ, the geometric structures which we have found
should emerge in some regime of large
radius and small fibers.

In the following subsections we will consider
torus and $K3$ fibrations in turn.

\subsection{Torus Fibrations and Type-II/M-theory Duality}

In the usual equivalence between M-theory and Type IIA
string theory, one employs simple Kaluza-Klein
reduction to the fields.
In the reduction from M-theory on $T^4$
to IIA on a three-torus, the metric
field of a four-torus gives a metric on the three-torus,
a Ramond-Ramond gauge field, and a scalar.
In our situation, if we take $X$ to be fibered by $C \cong T^4$,
then the Ramond-Ramond gauge field will
be varying and generically produce a non-zero field
strength.  This leads to the generation of a superpotential,
for an $\CN=1$ theory\footnote{If the fibration of $T^4$
over $T^3$ is not changing, then
the dilaton is constant and
$X \cong CY \times S^1,$
so we recover $\CN=2$ supersymmetry.} \cite{Gukov, TV, Mayr}.

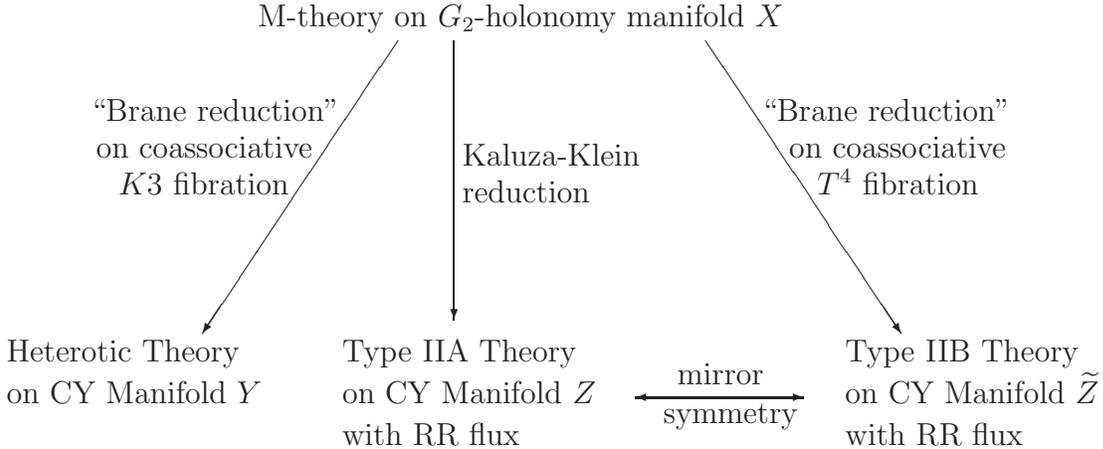
\begin{figure}
\setlength{\unitlength}{0.9em}
\begin{center}
\begin{picture}(25,15)

\put(2,15){M-theory on $G_2$-holonomy manifold $X$}

\put(7,14.5){\vector(-2,-3){7}}
\put(-4,11.6){``Brane reduction''}
\put(-3.8,10.3){on coassociative}
\put(-3,9){$K3$ fibration}

\put(9,14.5){\vector(0,-1){10}}
\put(9.3,10){Kaluza-Klein}
\put(9.3,8.7){reduction}

\put(18,14.5){\vector(2,-3){7}}
\put(20,11.6){``Brane reduction''}
\put(21,10.3){on coassociative}
\put(22,9){$T^4$ fibration}

\put(-7,3){Heterotic Theory}
\put(-7,1.5){on CY Manifold $Y$}

\put(5,3){Type IIA Theory}
\put(5,1.5){on CY Manifold $Z$}
\put(5,0){with RR flux}

\put(23,3){Type IIB Theory}
\put(23,1.5){on CY Manifold $\widetilde{Z}$}
\put(23,0){with RR flux}

\put(18.5,1.7){\vector(1,0){3}}\put(18.5,1.7){\vector(-1,0){3}}
\put(17,2.2){mirror}\put(16.5,0.8){symmetry}

\end{picture}\end{center}
\caption{Various dualities between M-theory on manifold $X$
with $G_2$ holonomy and string theory on Calabi-Yau manifolds.}
\label{figdual}
\end{figure}

%

However, in the reasoning of Sec. \ref{mfib},
we have done something {\sl different}
to arrive at a Calabi-Yau manifold starting from
$X,$ via self-dual forms.  What's the relation?
By analogy with the mirror symmetry argument of \cite{SYZ},
we should be performing not the Kaluza-Klein reduction,
but its {\sl mirror}.
Therefore, we should arrive at a
Type-IIB theory on our Calabi-Yau three-fold, and that
IIB theory should have a {\sl dual} torus fibration --- see
Fig. \ref{figdual}.
(Generalizations of the SYZ conjecture to $G_2$
manifolds with torus fibrations have been
made in \cite{BA} and \cite{LeeLeung}.)

To show this is true,
we isolate a vector $v$ in the direction of the M-theory circle.
Then, we use the $G_2$ structure to find a
perfect pairing between $\R^4/\R v$ and
the self-dual directions in $\R^4,$ where $\R^4$ is the
tangent space to the coassociative fiber.
But note that the $G_2$ three-form $\Phi_0$ on $\R^7$
already gives an isomorphism
$\Lambda^2_+\R^4 \cong (\R^4)^\perp,$
so we need only find, given a
vector $v,$ a pairing between vectors $w \in
\R^4/\R v$ and vectors $n$ normal to the four-plane.
This pairing is simply
\begin{equation}
\Phi_0(v,w,n),
\end{equation}
and is perfect.\footnote{For example, if we write
\begin{equation}
\Phi_0 = e_{125} + e_{345} + e_{136} - e_{246} + e_{147}
+ e_{237} + e_{567}
\end{equation}
and $v = e_1,$ then the duality pairs $e_2 \leftrightarrow
e_5,$ as
$\iota_v \Phi_0 = e_{25} + e_{36} + e_{47},$ etc.}

Let us call the Calabi-Yau formed from the moduli
space of coassociative torus fibers the ``brane reduction''
of the $G_2$ manifold $X,$
as opposed to the Kaluza-Klein reduction.
What we have argued then is that
{\sl brane reduction is mirror to Kaluza-Klein reduction.}

Another piece of evidence for this correspondence
can be found by studying the effect of deforming the
$C$-field in M-theory.  Such a deformation doesn't
affect the geometry of the Kaluza-Klein reduction,
but it does alter the $B$-field, hence the complexified
K\"ahler form.  How does it affect the brane reduction?
It should enter the equations for the two-form gauge field
on the five-brane, hence change the self-duality
condition.  This alters the pairing between base
and fiber directions for the brane reduction, hence
changes the complex structure of the Calabi-Yau.
We therefore see once again the mirror relation
between brane and Kaluza-Klein reduction.

As described in Remark \ref{otherconjs},
one would naturally conjecture that the Kaluza-Klein
lift of a brane-reduction would yield a mirror
$G_2$ manifold, in the sense of Shatashvili-Vafa
\cite{SV}.  In fact, B. Acharya informs us that he
has constructed $G_2$ orbifolds with $T^4$
and $T^3$ fibrations (and discrete torsion),
and finds that dualizing along $T^3$ {\sl or}
along $T^4$ fibers produces $G_2$ manifolds with
the same values of $b_2 + b_3,$ as required by \cite{SV}.
We thank him for informing us of these interesting
examples, which suggest that mirror Calabi-Yau manifolds
correspond to mirror $G_2$ manifolds.

\subsection{$K3$ Fibrations and Heterotic/M-theory Duality}

Consider heterotic string theory on a Calabi-Yau three-fold $Y$.
Following \cite{SYZ}, we view $Y$ as a fibration
by special Lagrangian tori over a 3-dimensional base ($={S}^3$).
Since heterotic string on $T^3$ is dual to M-theory
on $K3$, we can apply this duality fiberwise, and conclude that
heterotic string on $Y$ should be dual to M-theory on
a $G_2$-holonomy
manifold $X$, which in turn can be viewed as a fibration by (possibly
singular) $K3$ fibers.

Identifying BPS domain walls in
M-theory and in heterotic string, we can obtain a relation between
Betti numbers of $X$ and $Y$, assuming that both manifolds
are smooth \cite{PT}:
\begin{equation}
b_3 (X) = 2 h^{2,1} (Y) + 1
\end{equation}
Moreover Betti numbers of Calabi-Yau space $Y$ must obey $h^{2,1}=h^{1,1}$.
Finally, we could also look at the effective theory in M-theory
on $X$ and in heterotic theory on $Y$. The spectra of light particles
should match. In particular, we should expect the matching of
numbers of chiral/vector multiplets in dual descriptions.

In the simple case (when singularities in the K3 fiber can be
resolved or deformed), both $X$ and $Y$ are smooth manifolds.
It follows that the gauge group in the effective four-dimensional
theory is abelian, typically of rank (notice an obvious mistake
in \cite{PT}):
\begin{equation}
14 = {\rm ~number ~of~}\CN\!=\!1\,{\rm ~vector~ multiplets}
\end{equation}
This is the rank of the gauge group in heterotic theory broken
to a subgroup by Wilson lines, which can be continuosly
connected
to a 
trivial Wilson line, {\it i.e.}
we are on the main branch of the
moduli space
corresponding to the so-called {\it standard embedding}.

On the other hand, in M-theory on $G_2$ manifold $X$, vector fields
come from KK modes of the $C$-field. So, there are $b_2(X)$ of them.
Therefore, identifying $\mathcal{N}=1$ vector multiplets in the low-energy
effective theory, we seem to find a peculiar condition:
\begin{equation}
b_2 (X) \le 16
\end{equation}
for all $G_2$-holonomy manifolds (with generically non-singular
$K3$ fibers) that have heterotic duals.
Of course, we assume that the
heterotic dual is purely geometrical,
i.e. there are no space-filling five-branes.

In general, one needs space-filling branes to cancel anomalies.
For example, in F-theory
on a Calabi-Yau
four-fold these are D3-branes, needed to
cancel the
$\chi/24$ tadpole of the F-theory compactification.
On the other hand, in heterotic string theory on a Calabi-Yau space $Y$
these are five-branes wrapped on holomorphic curves inside $Y$.
Via duality to M-theory these space-filling branes become
certain singularities of the coassociative fibration (such
that the whole $G_2$ manifold may still be non-singular).


Summarizing, we argued that the Calabi-Yau dual of M-theory
on $X$ is a heterotic compactification
on the moduli space ${\cal M}_{\rm coassoc}$
of coassociative cycles in the deformation class of $C.$
There are a few important remarks in place here:

\begin{remark}
Note that the metric on the torus part of the
heterotic compactification is changing.
This scenario is somewhat similar to the stringy
cosmic string, in which the modulus of the
compactification has spatial dependence.
In that case, the equations of motion ensured
holomorphicity of the total space.
We would hope to find that the $G_2$ holonomy condition
is related to the equations of motion for the stringy
cosmic string in this generalized setting \cite{GSVY}.
\end{remark}

\begin{remark}
By analogy to the Calabi-Yau case, where a multiple of
the holomorphic form becomes an integral form near
the large radius limit,\footnote{This is
pointed out in \cite{OYZ}, for example.} one
anticipates that near some limit in $G_2$ moduli space,
the self-dual forms can be represented by integral
forms, and the left- and right-moving spaces compactify
separately (i.e. the lattice is compatible with the left/right split).
The limiting Calabi-Yau, then, would look like
the quotient of $T^*\msp$ by a lattice.
This is the $G_2$ analogue of the fact that
there is no quantum correction to the complex
structure of the special Lagrangian
fibration near the large radius limit.
Perhaps one could then use a Gauss-Manin connection
to follow the Calabi-Yau manifold away from this limit.
\end{remark}

%

We now highlight some features
of ${\cal M}_{\rm coassoc}$
which make it a possible candidate for the base
of a Calabi-Yau with torus fibration.
The moduli space $\msp$ has a natural metric on it,
given by the inner product of anti-self-dual forms:
\begin{equation}
g(V_1,V_2) = - \int_C \theta_1 \wedge \theta_2
\end{equation}
where $\theta_i$ are the anti-self-dual two-forms
corresponding to the tangent vectors $V_i \in T\msp\vert_C.$
We also have a correspondence
between moduli (harmonic self-dual forms) and
left-moving field strength directions (self-dual
cohomology classes).  This defines an almost complex
structure on the left-moving target space.
In addition, $\msp$ has a natural three-form $\Omega$
on it, defined as follows.  Let $V_1,V_2,V_3$ be three
vectors in $T\msp\vert_C,$ and let $v_1,v_2,
v_3$ be the corresponding normal vectors to $C$ in $X.$
We define
\begin{equation}
\Omega(V_1,V_2,V_3)\vert_C =
\int_C \Phi(v_1,v_2,v_3) dV,
\end{equation}
where $\Phi$ is the three-form defining the $G_2$ structure.
When $b_2^+(C) = 3,$ this is a top-dimensional form.
We expect, by analogy with the special Lagrangian
D-brane case, that this three-form gets complexified,
to define a holomorphic
three-form on the left-moving
part of the five-brane moduli space.
For more about geometric structures on $\msp$ see
\cite{LeeLeung}.


\subsection{Mirror Symmetry as Fourier-Mukai Transform}
\label{fiberwise}

Using various dualities between
string theory on Calabi-Yau manifolds and
M-theory on $G_2$-holonomy manifolds,
we argued that such $G_2$-holonomy manifolds
should be fibered by coassociative tori or $K3$ surfaces.
In this subsection, we come to the same conclusion
using only string dualities and interpreting mirror
symmetry for $G_2$ manifolds as a Fourier-Mukai transform
on the coassociative fiberes.

For concreteness, consider a $K3$ fibration,
and let $B$ be the base of this fibration:
\begin{equation}
\pi \colon X  \to  B.
\end{equation}
Let us analyze in more detail the structure of this fibration.
One natural question about this fibration by $K3$ surfaces is about
the geometric meaning of the Fourier-Mukai transform acting on
each fiber. By analogy with the SYZ conjecture \cite{SYZ},
it is natural to expect that this transformation
corresponds to a symmetry of the full quantum string theory
on $X$, viz. to T-duality or mirror symmetry \cite{SV}.
In order to follow the arguments of \cite{SYZ}
in the $G_2$ case, let us go to Type-IIA theory on $X$.

We can take a D0-brane on $X$, with moduli space equal to $X$.
Locally, we can identify the moduli with $(p,q)$, where $q$ is
the position of the D0-brane on $B$, and $p$ is its position in
a $K3$ fiber. On the 7-manifold $X$ we can describe a D0-brane
as a (non-holomorphic) skyscraper sheaf, $\mathcal{E}$,
supported at $(p,q)$.  Now
since everything
is going to happen in the fiber, we can think of $\mathcal{E}_p$
as a (holomorphic) coherent sheaf on the $K3$ at $q.$
Let $v( \mathcal{E}_p)$ be the corresponding Mukai vector:
\begin{equation}
v(\mathcal{E}_p) = \hbox{D-brane charge}
= ch (\mathcal{E}_p) \sqrt{\hat A(K3)}.
\end{equation}
In particular, for a D0-brane we have $v=(0,0,-1)$.

Naively, one might expect that via Fourier-Mukai transform
a D0-brane becomes a D4-brane wrapped on the entire $K3$ fiber.
It should be also $1/2$-BPS, so immediately we infer that $K3$
fibers should be volume minimizing, i.e. coassociative
submanifolds inside $X$.

As the number of
deformations of coassociative $K3$'s in $X$
equals $b_2^+(K3)=3,$ it is
natural to identify position on the base, $q \in B$, with the
local coordinate on this moduli space. But this clearly cannot
be the full story since the moduli space of the original D0-brane
was 7-dimensional (a copy of $X$), and the same should be true for
the dual D4-brane.

The solution is that after we make a Fourier-Mukai
transform we obtain the Mukai vector
\begin{equation}
v_{dual} = (1,0,0).
\end{equation}
This is not the right charge vector for a D4-brane on $K3$. Since
$p_1(K3)=48$, the latter would be $v=(1,0,-1)$. So, after performing
Fourier-Mukai transform we actually get a bound state of D4-brane
and a D0-brane! It has the right charge vector, $v=(1,0,0)$, and
the right dimension of the moduli space. In fact, according to
Mukai, the real dimension of the moduli space of a sheaf
$\mathcal{E}$ with Mukai vector $v=(r,l,s)$ is
\begin{equation}
dim ( M ( \mathcal{E} )) = 2 l^2 - 4 r s + 4.
\end{equation}
which is equal to 4
when $v = (1,0,0).$  So, the total dimensional of the
moduli space of the dual D0/D4 bound state is indeed equal to 7,
as expected, in complete analogy with the SYZ case.
Note, that instead of $K3$ we could take
$T^4$ as a coassociative fiber.
In this case, the story is much easier: there is no induced
D0-brane charge on the dual D4-brane, after we make four
T-dualities along the $T^4$. In this case again one has
$b_3^- (T^4)=3$ for the number of deformations of coassociative
$T^4$ cycle, and $b_1 (T^4)=4$ for the number of moduli associated
with flat connections. Hence, the total dimension of
the moduli space of dual D4-brane is equal to 7,
which is the right dimension to describe mirror $G_2$ manifold.
This case was already studied in \cite{Bobby}.

\subsubsection*{\rm{\em Rigidity of the Base ?}}

In both $K3$ and $T^4$ fibrations, we could take an appropriate
D7-brane wrapped on the entire $X$, and conclude that $B$ is
itself a supersymmetric 3-cycle in $X$ --- an associative cycle.
Indeed, dualizing the D7-brane along the fibers we find a D3-brane
wrapped around the base $B$. Since the moduli spaces of these
two D-branes should be the same, one might expect that both are rigid.
In fact, the D7-brane does not have any geometric deformations.
Furthermore, $\mathrm{Hol} (X) = G_2$ implies $b_1 (X) =0$,
which means that the space of flat $U(1)$ gauge connections
on the D7-brane is also zero-dimensional. However, a complete
answer to this question should involve a more careful analysis
of the gauge bundle on the D7-brane, and it would interesting
to study it further both from physics and mathematics points of view.



\section{A $K3$ Fibration}
\label{k3fib}

\subsection{Idea and Basic Set-Up}

Imagine a $G_2$ manifold which is a $K3$
fibration over a base $S^3,$ with a discriminant
locus $\Delta,$ which we assume to be a closed
manifold of codimension two --- a knot or link.
If we consider the case of a non-satellite
knot, then by Thurston's theorem
there exists a hyperbolic metric
on the complement $S^3 \setminus \Delta.$
In this section, we use this reasoning to look for
a $G_2$ structure on a $K3$ fibration
$X$ over a non-compact hyperbolic
manifold.\footnote{This construction works for
$T^4$ fibrations too, if in the following
we simply replace
$SO(3,19)$ and its maximal compact
subgroup by $SO(3,3)$ with its corresponding
subgroup.}
For simplicitly, we take the contractible
hyperbolic space $B = SO(3,1)/SO(3)$
for our base, with its hyperbolic metric
$g_B,$ left-invariant under the action of
$SO(3,1).$
Thus, $X = B \times K3$ as a differentiable manifold.
We write $$\pi: X\rightarrow B$$
for the projection to base.
Note that at a point $p \in X$ the vertical vectors are
defined as the kernel of $\pi$ and span a sub-bundle
$T_VX$ of $TX,$ but there is no canonical notion
of horizontal vectors until we have a {\sl connection,}
i.e. a choice of ``horizontal'' subbundle $T_HX$ of $TX.$
Such a choice allows us to decompose $TX$ as
$TX = T_HX \oplus T_VX,$ and we write $P_H$ and $P_V$
for the corresonding projection operators.
We will discuss such a choice in
section \ref{metsec}.

Choose over a point $b\in B$ a marking for the $K3$
fiber.  Recall from Ref.~\cite{Kob}
that the moduli space of
Einstein metrics on a marked $K3$ manifold with
unit volume is isomorphic to
$${\cal M}_{K3} = SO(3,19)/[SO(3)\times SO(19)].$$
(Quotienting on the left by $SO(3,19;\Z)$
removes the choice of marking.)
Since we will
wish to fiber $X$ with K\"ahler-Einstein
$K3$'s, we will employ a map from $B$ to ${\cal M}_{K3}.$
Let ${\tau}: B\rightarrow {\cal M}_{K3}.$
The next section now shows that we can choose
a family of Ricci-flat metrics for $K3$'s
over $B$ realizing the
family defined by $\tau,$
and that we have a natural connection.

\subsection{The fiberwise metric and connection}
\label{metsec}

Let $RicMet$ be the space of unit-volume
Ricci-flat metrics
on a fixed differentiable $K3$ manifold.  On $RicMet\times K3$
we have a natural fiberwise metric: at
$(p,q)$ we have the metric $p$ defines at $q.$
Then the
group of diffeomorphisms $Diff$
acts on $RicMet$ on the
right via pull-back, and on $K3$ on the left.
The fiber product
$RicMet \times_{Diff} K3$ inherits the
fiberwise metric, and defines a
universal family of Ricci-flat $K3$ manifolds over
${\cal M}_{K3} = RicMet/Diff.$
This is universal in the sense that any
other family of Ricci-flat $K3$ metrics
can be mapped to the constructed
family by a canonical family of diffeomorphisms,
with the fiberwise metric defined by pull-back.

We now can assume that $X$ has been chosen to
realize the map $\tau,$ i.e. that
$X\rightarrow B$ is a family
of Ricci-flat $K3$'s such that
the equivalence class of
the metric on $\pi^{-1}(b)$ equals $\tau(b).$
We now construct a canonical connection
on $X.$
We need to define the horizontal sub-bundle of $TX.$

We will define the horizontal part of a vector $V$ in
$X$ at $q,$ a point in the fiber over
$b\in B.$
Let $U\subset B$ be a neighborhood containing
$b$ such that $\pi^{-1}(U) \cong U \times K3.$
Using a trivialization for the fiber bundle
$X\rightarrow B,$ we may
assume we have a family of metrics on a {\sl fixed}
K3.
Let $t$ parametrize a path $\gamma(t)$ in the
base through $b$ such that $\dot{\gamma}(0) = \pi_*V.$
Then $g_t$ (the metric over $\gamma(t)$)
defines a family of metrics, and we look for
a family of diffeomorphisms $f_t : K3 \rightarrow K3$
such that
\begin{equation}\label{gauge}
f_0 = {\rm id},\qquad
\frac{d}{dt}(f_t^*g_t) \perp \hbox{
gauge orbit of diffeos}.
\end{equation}
Here perpendicularity is in the space of
metrics, which is equipped with the natural
metric on symmetric two-tensors.
As we will show, this uniquely determines
$f_t.$  We therefore get a curve $\Gamma(t)$
passing through $q$ defined by
$\Gamma(t) = (\gamma(t),f_t(q)),$ and we define the
horizontal component of $V$ to be
$$V_H \equiv \frac{d}{dt}\Gamma(t)\vert_{t=0}.$$

\begin{lemma}
The conditions in (\ref{gauge}) uniquely determine
$f_t.$
\end{lemma}
\proof
In fact we only need the first derivative of
$f_t$ at $t=0,$ which is defined by a vector field $\hat{\xi}.$
Using the
metric, we can equate this
with a one form $\xi = \xi_\mu dx^\mu,$
written in local coordinates $x^\mu.$
Let $\eta = \frac{d}{dt}g_t\vert_{t=0}.$
Then
$$A_{\mu\nu} \equiv
\frac{d}{dt}(f_t^* g_t)_{\mu\nu}\vert_{t=0}=
\eta_{\mu\nu} + \nabla_\mu \xi_{\nu} +
\nabla_\nu \xi_\nu.$$
Now perpendicularity of $A_{\mu\nu}$
to the gauge orbit
under diffeomorphisms means
\begin{equation}
\label{gaugecond}
\nabla^\mu A_{\mu\nu} = 0.
\end{equation}
Imposing this condition
leads to the equation
\begin{equation}
\label{nabeq}
\nabla^{\mu}\nabla_\mu \xi_{\nu} +
\nabla^\mu \nabla_\nu \xi_\nu = -\nabla^\mu \eta_{\mu\nu}
\equiv - B_\nu,
\end{equation}
where we have defined the one-form
$B_\nu =  \nabla^\mu \eta_{\mu\nu}.$
Now note
$$\nabla^\mu\nabla_\nu\xi_\mu
= [\nabla^\mu,\nabla_\nu]\xi_\mu + \nabla_\nu\nabla^\mu
\xi_\nu = R_\nu{}^\alpha\xi_{\alpha}+ \nabla_\nu\nabla^\mu
\xi_\nu = \nabla_\nu\nabla^\mu
\xi_\nu, $$
where we have used Ricci flatness.
Therefore, (\ref{nabeq})
becomes
$$\nabla^{\mu}\nabla_\mu \xi_{\nu} +
\nabla_\nu \nabla^\nu \xi_\nu = - B_\nu,$$
or
$$d^\dagger d\xi + 2 dd^\dagger\xi = B.$$
Using Hodge decomposition and the fact that
$K3$ is simply-connected, we can write
$\xi = dh + d^\dagger k$ for some
function $h$ and two-form $k,$ and we may
choose $h$ and $k$ perpendicular to the
kernel of $d$ and $d^\dagger,$ respectively.
Likewise, $B$ has a decomposition as $dH + d^\dagger K,$
with $H$ and $K$ chosen similarly.
Therefore
$d^\dagger d d^\dagger k + 2 d d^\dagger d h =
dH + d^\dagger K,$
and we have
$$k = \frac{1}{d d^\dagger} K, \qquad
h = \frac{1}{2 d^\dagger d} H.$$
This determines $\xi,$ thus $\frac{df_t}{dt}\vert_{t=0},$
uniquely.
$\blacksquare$

The set of vectors $V_H$ spans
a sub-bundle $T_H \subset TX.$
Further, we have a splitting $TX = T_H \oplus T_V$
defined on $U,$
where $T_V$ is the vertical sub-bundle.
The definition of the splitting
is independent of the trivialization, so
we have shown:
\begin{corollary}
The sub-bundle $T_H$ is well-defined on
all of $X$ and defines a connection or splitting,
$TX = T_H \oplus T_V.$
\end{corollary}

\subsection{Defining the Three-form, $\Phi$}

{}From the previous construction,
we can now form a natural metric on $X$
as follows.  Let $p \in X,$
$b = \pi(p)$ and $\mu = {\tau}(b),$
with $g^\mu_K$ the metric
defined by $\mu\in {\cal M}_{K3}.$
Let $U,V \in T_pX.$  Then
we can define
\begin{equation}
\label{metric}
g(U,V) = g_B(\pi_*U,\pi_*V)\vert_{b}
+ g^\mu_K(P_V U,P_V V)\vert_{p}
\end{equation}
(one could also multiply the two terms by
positive functions of $X$),
where $P_V$ is the vertical projection.
In particular, horizontal and vertical directions
are declared to be perpendicular.

Note that both $B$ and ${\cal M}_{K3}$ are homogeneous
spaces $G/H.$  Then $H$ acts on the tangent space
of $G/H,$ since the stabilizer of the transitive
left $G$ action at $[g]$ is $H_g \equiv
gHg^{-1} \cong H.$  This
is obvious, since $(ghg^{-1})[g] = [gh] = [g],$
where $[g]$ denotes the coset $gH.$

Now let $\widetilde{\tau}:
SO(3,1)\rightarrow SO(3,19)$
be a group homomorphism such that
$$\widetilde{\tau}(SO(3))\subset SO(3)\times SO(19)$$
and
$$p_1\circ \widetilde{\tau} \hbox{
is an isomorphism,}$$
where $p_1$ is the projection to the first
factor.
Now $\widetilde{\tau}$ induces a map
$$\tau: B \rightarrow {\cal M}_{K3},$$
equivariant in the following
sense.  Let $c \in SO(3),$ which
acts on $TB.$  Let $C =
\widetilde{\tau}(c) \in SO(3)\times SO(19)$
acting on $T{\cal M}_{K3}.$
Then
\begin{equation}
\tau_* \circ c = C \circ \tau_*,
\end{equation}
where $\tau_*$ is the push-forward of tangent
vectors.
Namely, we have equivariance under corresponding
$SO(3)$ actions.\footnote{This
construction was meant
to mimick the notion of holomorphicity for an
an elliptic fibration, which can be written
as the equivariant property
$\tau_* \circ j = J \circ \tau_*,$ where $j$ and $J$
are complex structures on the
base of the fibration and the
moduli of elliptic curves, respectively.
We thus think of a complex
structure as
the $\pi/4$ element of an $S^1 = SO(2)$ action.
In our case, $SO(3)$ is the relevant group.}

We now try to construct a positive
$G_2$ calibration $\Phi$ on $X$,
so that $\Phi$ is
a nowhere-vanishing, closed three-form
form which at every point $p\in X$ lies in the
$GL(T_p X)$ orbit of the standard associative
form $\Phi_0$ (encoding the structure constants
of multiplication on the imaginary octonians).
Recall that the construction of $\Phi_0$ involves
writing $\R^7$ as ${\rm Im}\HH\oplus \HH,$ then identifying
for each orthonormal basis element $e_i$ in ${\rm Im}\HH$
a self-dual form $\alpha_i$ in $\Lambda^2\HH$
which encodes multiplication (in $\HH$) by $e_i.$
For example, $\alpha_1 = (e_4 \wedge e_5 + e_6 \wedge e_7).$
Then $\Phi_0 = e_1 \wedge e_2 \wedge e_3 + \sum_{i = 1}^{3}
e_i \wedge \alpha_i.$  A $G_2$ form $\phi$
defines a metric $g$ by $g(u,v) dV =
\iota_u \phi \wedge \iota_v \phi \wedge \phi,$
where $dV = \sqrt{{\rm det}g}\, e_1\wedge ...\wedge e_7.$

To construct such a $\Phi$ then, we must relate the
tangential directions on $B$ to self-dual forms
on the fiber.  (The metric on $B$ allows us to
equate tangent vectors and one-forms.)
Recall that $TB\vert_{[g]} \cong
\g/\h_g,$ where $\g = so(3,1),$
$\h = so(3),$ and we have
defined $\h_g = g\h g^{-1} =
{\rm Ad}_g \h$ (independent of
the representative of the coset $[g]$).
Our key construction will be the simple observation
that
\begin{equation}
\label{geqs}
\g/\h_g \cong \h_g^\perp \cong \h_g
\end{equation}
as vector spaces, where
the first equivalence comes from the metric on $\g$
and the second comes from the (pseudo-)symplectic structure
on $\g = so(3,1) \cong sl(2,\C) \cong \C^3,$
by which $\h$ is a Lagrangian
subspace.\footnote{Explicitly, at the identity coset,
we identify $so(3,1)=sl(2,\C)$
with traceless, $2\times 2$ complex matrices
with the indefinite metric
$\langle A,B\rangle = -\frac{1}{2}{\rm Tr}(AB),$
then $so(3)=su(2)$ are the anti-hermitian ones,
the map $\g/\h\rightarrow \h^\perp$ sends
$A \mapsto (A - A^\dagger)/2.$
The involution
from $\h^\perp \leftrightarrow \h$ is just
$A \mapsto iA^\dagger,$ or $\sqrt{-1}$ times
the Cartan involution.
Put physically,
the projection to $\h^\perp$ eliminates rotational
pieces of an infinitesimal Lorentz transformation,
and then the correspondence $\h^\perp\cong\h$
associates to a pure boost in some direction a
rotation in the same direction.}
This allows us to associate to $V \in T_gB$
an element $C_V \in so(3).$
Next we shall get from $C_V$ a self-dual form on the fiber.
Recall now that ${\cal M}_{K3}$ is a Grassmannian
of positive, oriented
three-planes in $\R^{3,19},$
which is interpreted as the plane
of self-dual harmonic two-forms
inside $H^2(K3).$  The first factor $SO(3)\subset
SO(3)\times SO(19)$ acts on the three-plane.
An element of $SO(3)$ singles out a direction
defined by its zero eigenspace (the axis defining
the rotation in three-space).
Therefore, from $V$ at $b$ we form $C_V$ which
maps by $p_1\circ\widetilde{\tau}$ to $so(3)$ inside
the stabilizer of the three-plane of
self-dual forms on $\pi^{-1}(b).$
Let $\theta_V$ be a generator of
${\rm Ker}[p_1(\widetilde{\tau}(C_V))],$
defined up to sign, and normalized so that
$\int (\theta_V)^2/2 = \vert V\vert^2.$
(The signs can be chosen consistent
with the orientation.\footnote{If
we write an element $C_V$ of $so(3)$ as $aX + bY + cZ,$
then $(a,b,c)$ defines the axis of rotation in
three-space.  After equating
the three-plane $\R^3$ isometrically
with $so(3),$ then $\theta_V$ is simply $(a,b,c).$
That is, $End(E) \cong E$ for oriented,
three-dimensional metric
vector spaces.})
This is our sought-after self-dual two-form, and
we have found a
map
$$\theta: T_bB \rightarrow H^2_+(\pi^{-1}(b)),$$
defined up to sign.
In short, we have described another sequence of
isomorphisms
\begin{equation}
\label{moreeqs}
\h_g \cong so(3)
\begin{matrix}
{}_{p_1\circ\widetilde{\tau}} \cr \cong \cr {}
\end{matrix}
so(H^2_+(K3)) \cong H^2_+(K3).
\end{equation}
$\theta$ is the composition of the isomorphisms
in (\ref{geqs}) and (\ref{moreeqs}).
Note that the forms in $H^2_+$ are harmonic, so
no choice of representative class is necessary.

Now let $p\in X$ with $\pi(p) = [g] \in B.$
Let $\{e_i\}$ be
the pull-back of an orthonormal frame for $T^*_{[g]}B \cong
\g/\h_g,$ and let $\{\theta_i\}$ be the corresponding
self-dual two-forms on $\pi^{-1}(b)\cong K3.$
We construct
\begin{equation}
\label{g2formframe}
\Phi = Vol_B + \sum_{i}e_i \wedge \theta_i,
\end{equation}
where $Vol_B$ is the volume form on $B$
and the pull-backs of the various forms to
$X$ are understood.  More invariantly, we can
write
\begin{equation}
\label{g2form}
\Phi(U,V,W) = Vol_B(\pi_*U,\pi_*V,\pi_*W) +
\left( \theta_{\pi_*U}(P_V V,P_V W) + {\rm cyclic}\right).
\end{equation}
Note that as $\widetilde{\tau}$ is an inclusion
of groups, then since the $e_i$ are orthogonal,
the unit-norm self-dual forms $\theta_i$ are
mutually orthogonal,
and the metric defined from $\Phi$ agrees with (\ref{metric}).



\subsection{Explicit formulas}
\label{explicit}

The constructions above can be made explicit.
Let $\eta_{\mu\nu}$ be a
metric deformation of K3,
perpendicular to diffeomorphisms.

\begin{lemma}
$\eta$ is locally volume-preserving.
\end{lemma}
\proof
The globally volume-preserving
deformation of the metric $\eta$ induces
a change of the Riemann tensor.
Imposing the gauge condition (\ref{gaugecond})
means $\nabla^{\mu}\eta_{\mu\alpha} = 0.$
Working in Riemann normal coordinates
and
using the fact that the Ricci curvature is
zero ($R^{\mu}{}_{\alpha\mu\beta}=0$),
one can compute that vanishing of the infintesimal
variation of the Ricci curvature is equivalent to
$$\nabla^\mu\nabla_\mu \eta_{\alpha\beta}+
2R_{\alpha}{}^\mu{}_\beta{}^\nu\eta_{\mu\nu} +
\nabla_\alpha\nabla_\beta\eta^\mu{}_\mu = 0.$$
Multiplying by $g^{\alpha\beta}$ and summing,
using
the fact that the metric is covariantly constant
and Ricci flat,
gives
$$2\nabla^\mu\nabla_\mu \eta^\alpha{}_\alpha = 0.$$
Now $\triangle (\eta^{\alpha}{}_\alpha) = 0$
means that $(\eta^{\alpha}{}_\alpha)$ is a constant, $C,$
but since $\eta$ is globally volume-preserving,
$\int C = C\cdot Vol = 0,$ and so $\eta$ is pointwise
traceless.
$\blacksquare$

Now let $S^a,$ $a = 1,...,3,$ be an orthonormal
set of self-dual forms.  (In an orthonormal
frame, the $S^a$ are antisymmetric matrices
obeying the algebra of the
quaternions $i, j, k,$ and $\eta$ is traceless.)
Then we can define the
anti-self-dual forms $A^a$ as follows:
$$(A^a)_{\mu\nu} = \frac{1}{2}(S^a{}_{\mu\sigma}
\eta^{\sigma}{}_\nu + \eta_{\mu\sigma}S^a{}^\sigma{}_\nu).$$
Conversely, if we have a trio of anti-self-dual
forms $A^a$ we can define a metric
deformation
$$\eta_{\mu\nu} =
- \sum_{a} A^a {}_{\mu\sigma} S^a{}^\sigma{}_\nu.$$
These operations are indeed inverses of each other
when $\eta$ is traceless and symmetric.\footnote{This
depends on some
nice facts, including
the following identity.  Let $A$ be
a traceless, symmetric, four-by-four matrix
acting on the quaternions $\R^4.$  Let
$I,$ $J,$ $K$ be matrices
representing multiplication by $i,$ $j,$ $k.$ Then
$A = IAI + JAJ + KAK.$}

To define the self-dual two-form defined by a metric
deformation, we can do the following.
Fix at a point in K3 moduli space a four-plane $W$
in $\R^{3,19}\cong H^2(K3)$ with signature $(3,1)$ and
containing $H^2_+(K3).$  Then choose
$\widetilde{\tau}$ to be an isomorphism
$SO(3,1)\cong SO(W).$  Then at each point
$b\in B$ there is a unique anti-self-dual
form $\beta$ in $W$ perpendicular to the $S^a,$
so that $A^a = r^a \beta$ for all $a = 1,...,3.$
Now each tangent vector $V$ on $B$
defines a metric deformation $\eta,$
which defines a trio of anti-self-dual forms $A^a,$
which then define a single, self-dual form
$$\theta_V = \sum_a r^a S^a.$$
This is the isomorphism described in
(\ref{geqs}) and (\ref{moreeqs}).

\subsection{$d\Phi \neq 0$}
We have not been able to show that $\Phi$ is closed,
and in fact this seems unlikely, despite the fact
that the map $\tau$ was meant to mimick the
(more successful) stringy cosmic string
construction. However, we believe that this may lead to
weak holonomy $G_2,$ for which $d\Phi = \lambda *\Phi.$
Then, one would also hope that this construction
can be modified to produce a torsion-free $G_2$ structure,
hence a manifold with true $G_2$ holonomy.
This is currently under investigation.

Also, it is worth mentioning that
since our constructions are left-invariant under the
transitive action of $G$ on $B,$ the behavior of
$\Phi$ can be analyzed at a single point, e.g. the
identity coset.
Also, if $\widetilde{\tau}$ maps the discrete
subgroup $SO(3,1;\Z)$ to the
subgroup $SO(3,19;\Z)$ then this entire
construction will descend to the
finite-volume quotient by this discrete group.



\section{A Torus Fibration}
\label{torusfib}

\subsection{Outline of Hitchin's Construction}

Recently Hitchin has shown how certain functionals on
differential forms in six dimensions generate metrics
with $G_2$ and weak $SU(3)$ holonomy \cite{Hit,Hitchin}.
Here, we outline his construction and
use his result to construct new $G_2$ metrics.
The main point is to consider the Hamiltonian flow
of a volume functional on a symplectic space
of stable three- and four-forms on a
six-manifold.
When a group acts on the six-fold, the invariant
differential forms can restrict the infinite-dimensional
variational problem to a finite-dimensional
set of equations governing the evolution.
Including the ``time'' direction, one is able
to create a closed and co-closed $G_2$ three-form,
thus a metric of $G_2$ holonomy.

The key element in the construction is the following:

\begin{theorem}
\label{hitthm}
\cite{Hitchin}:
Let $M$ be a 6-manifold,
${\cal A} \in H^3(M, \mathbb{R})$
and ${\cal B} \in H^4(M, \mathbb{R})$
be fixed cohomology classes,
and let $(\rho, \sigma) \in {\cal A} \times {\cal B}$
be stable forms of positive type which evolve via Hamiltonian
flow of the functional:
\begin{equation}
H=V(\rho)-2 V(\sigma).
\end{equation}
Here, $V(\rho)$ and $V(\sigma)$ are suitable
volume forms (which we define below),
with $\phi$ their integrands:  $V = \int_M \phi.$
If for some $t=t_0,$ $\rho$ and $\sigma$
safisfy the compatibility conditions
$\omega \wedge \rho=0$ and $\phi(\rho)=2 \phi(\sigma)$
(where $\sigma=\omega^2/2$) then the three-form
\begin{equation}
\Phi=dt \wedge \omega+\rho,
\label{formgtw}
\end{equation}
defines a $G_2$ structure on $X = M \times (a,b)$
for some interval $(a,b)$.
\end{theorem}

The converse is also true \cite{Hitchin}.

Stable forms are defined in an earlier paper by Hitchin:

\begin{definition} \cite{Hit}:
Let $M$ be a manifold of real dimension $n$, and $V = TM$.
Then, the form $\rho \in \Lambda^p V^*$ is stable if it lies
in an open orbit of the (natural) $GL(V)$ action on $\Lambda^p V^*$.
\end{definition}

In other words, this means that all forms in the neighborhood
of $\rho$ are $GL(V)$-equivalent to $\rho$.
This definition is useful because it allows
one to define a volume.
For example, a symplectic form $\omega$ is stable if and only if
$\omega^{n/2} \ne 0$.

Relevant to our discussion are 3-forms and 4-forms
on a 6-manifold $M$. If these forms are stable, we
can define the corresponding volumes as follows.
Let's start with a stable 4-form
$$
\sigma \in \Lambda^4 V^* \cong
\Lambda^2 V \otimes \Lambda^6 V^*.
$$
Therefore, we find
$$
\sigma^3 \in \Lambda^6 V \otimes (\Lambda^6 V^*)^3
\cong (\Lambda^6 V^*)^2
$$
and
\begin{equation}
V(\sigma)=\int_M |\sigma^3|^{\frac{1}{2}}.
\label{defvos}
\end{equation}

In order to define the volume $V(\rho)$ for a 3-form
$\rho \in \Lambda^3 V^*$, one first defines a map
$$
K_{\rho} \colon V \to V \otimes \Lambda^6 V^*,
$$
such that for a vector $v \in V=TM$ it gives
\begin{equation}
K (v) = \imath (v) \rho \wedge \rho \in \Lambda^5 V^* \cong
V \otimes \Lambda^6 V^*.
\label{oprk}
\end{equation}
Hence, one can define
$$
tr (K^2) \in (\Lambda^6 V^*)^2.
$$
Since stable forms with stabilizer $SL(3,\mathbb{C})$
are characterised by $tr(K)^2 < 0$, following \cite{Hit},
we define
\begin{equation}
V(\rho)=\int_M |\sqrt{-tr K^2}|.
\label{defvor}
\end{equation}

The last fact used in the Hitchin's theorem is that
there is a natural symplectic structure on the space
$$
{\cal A} \times {\cal B}
\cong \Omega^3_{exact} (M) \times \Omega^4_{exact}.
$$
Explicitly, it can be written as
$$
\omega \left( (\rho_1, \sigma_1) , (\rho_2, \sigma_2) \right)
= \langle \rho_1, \sigma_2 \rangle - \langle \rho_2, \sigma_1 \rangle,
$$
where, in general, for
$\rho = d \beta \in \Omega^p_{exact} (M)$
and $\sigma = d \gamma \in \Omega^{n-p+1}_{exact} (M)$
one has a nondegenerate pairing
\begin{equation}
\langle \rho, \sigma \rangle = \int_M d \beta \wedge \gamma
= (-1)^p \int_M \beta \wedge d \gamma.
\end{equation}

Then, Hitchin shows that the first-order Hamiltonian flow equations
in the theorem quoted above are equivalent to the closure and
co-closure of the associative form $\Phi = dt \wedge \omega + \rho$:
$$
d \Phi =0, \quad d * \Phi =0.
$$

In order to construct the metric with $G_2$ holonomy from
the form $\Phi$ we should
take $v, ~w \in W,$ where $W = TX$ is the seven-dimensional
vector space and define a symmetric bilinear form on $W$
with values in $\Lambda^7 W^*$
by
\begin{equation}
B_{\Phi}=-{1 \over 6} \imath(v) \Phi \wedge \imath(w) \Phi \wedge \Phi.
\label{oprbb}
\end{equation}
This defines a
linear map $K_{\Phi}: W \to W^* \otimes \wedge^7 W^*.$
Then the $G_2$ holonomy metric can be written as \cite{Hit}
\begin{equation}
g_{\Phi}(v,w)=B_{\Phi}(v,w) (\det K_{\Phi})^{-{1 \over 9}}.
\label{hitmet}
\end{equation}

\subsection{Equations for the Metric}

Now let us consider an example of (non-compact) $G_2$ manifold,
with principal orbits
$$
M = {S}^3 \times T^3.
$$
We think of ${S}^3$ as a group manifold $SU(2)$.
The space $M$ appears as one of the examples in the recent
work of Cleyton and Swann \cite{Swann}, where they classified
principal orbits of cohomogeneity-one $G_2$ manifolds under
a compact, connected Lie group.

In order to construct differential forms $\rho$ and $\sigma$,
let us choose a basis of left-invariant one-forms on $SU(2)$:
\begin{eqnarray}
\Sigma_1 &=& \cos \psi d \theta
+ \sin \psi \sin \theta d \phi, \nn \\
\Sigma_2 &=& - \sin \psi d \theta
+ \cos \psi \sin \theta d \phi, \nn \\
\Sigma_3 &=& d \psi + \cos \theta d \phi.
\label{sutwoforms}
\end{eqnarray}
which enjoy the $su(2)$ algebra
\begin{equation*}
d \Sigma_a=-{1 \over 2} \epsilon_{abc} \Sigma_b \wedge \Sigma_c.
\label{sualg}
\end{equation*}

We also choose closed, but not exact one-forms $\alpha_i$,
which generate the $H^1(T^3)$ cohomology of the torus:
\begin{equation*}
\alpha_{1,2,3} \in H^1 (T^3) = \mathbb{R}^3.
\end{equation*}
Explicitly, if we define
\begin{equation*}
T^3 = \mathbb{R}^3 / \mathbb{Z}^{3},
\end{equation*}
where $\mathbb{R}^3$ is parametrized by affine coordinates
$u_1$, $u_2$, and $u_3$, we can write
\begin{equation*}
\alpha_i = d u_i, \quad i=1,2,3.
\end{equation*}

Now, we have to fix cohomology classes $\cal{A}$ and $\cal{B}$
in $H^3 (M)$ and in $H^4 (M)$, respectively.
The cohomology groups are non-trivial:
\begin{eqnarray}
H^3 (M; \mathbb{R}) & = & \mathbb{R} \oplus \mathbb{R}, \nn \\
H^4 (M; \mathbb{R}) & = & \mathbb{R} \oplus \mathbb{R} \oplus \mathbb{R},
\nn
\end{eqnarray}
so that our choice depends on five {\sl real}
parameters that we call
$m$, $n$, $k_1$, $k_2$, and $k_3$. As we will see in a moment,
the construction depends only on
the parameters $m$ and $n$, which
determine the class ${\cal A} \in H^3 (M; \mathbb{R})$.
Specifically, $m$ corresponds to the class $[T^3]$
and $n$ corresponds to the class $[{S}^3]$.

Now we can write the
$SU(2)$-invariant 3-form $\rho \in \cal{A}$ as
\begin{equation*}
\rho = n \Sigma_1 \Sigma_2 \Sigma_3 - m \alpha_1 \alpha_2 \alpha_3
+ x_1 d ( \Sigma_1 \alpha_1) + x_2 d ( \Sigma_2 \alpha_2)
+ x_3 d ( \Sigma_3 \alpha_3).
\end{equation*}
Here, $x_i (t)$ are functions of the extra variable $t$
that describe variation of the 3-form $\rho$ within a given
cohomology class (determined by $m$ and $n$).
The radial direction $t$ is going to play the role of
time variable for the Hamiltonian evolution.
Clearly, the form $\rho$ is colsed.

Similarly, we can write a natural 4-form:
\begin{eqnarray}
\sigma =
k_1 \Sigma_1 \Sigma_2 \Sigma_3 \alpha_1
+ k_2 \Sigma_1 \Sigma_2 \Sigma_3 \alpha_2
+ k_3 \Sigma_1 \Sigma_2 \Sigma_3 \alpha_3 \nn \\
+ y_1 \Sigma_2 \alpha_2 \Sigma_3 \alpha_3
+ y_2 \Sigma_3 \alpha_3 \Sigma_1 \alpha_1
+ y_3 \Sigma_1 \alpha_1 \Sigma_2 \alpha_2.
\nn\end{eqnarray}
The first line in this expression is cohomologically non-trivial,
whereas the second line contains three exact terms. Indeed,
\begin{equation*}
\Sigma_2 \alpha_2 \Sigma_3 \alpha_3 = d (\Sigma_1) \alpha_2 \alpha_3
= d (\Sigma_1 \alpha_2 \alpha_3 ),
\end{equation*}
and similarly for other terms.
Therefore, both $\rho$ and $\sigma$ are closed forms.

In order to see that parameters $k_{1,2,3}$ are irrelevant,
let us evaluate the volume corresponding to the form $\sigma$:
\begin{equation*}
V^2 (\sigma) = y_1 y_2 y_3.
\end{equation*}
Since it does not depend on the choice of the cohomology
class, in what follows we set $k_1 = k_2 = k_3 =0$.
Hence, the $SU(2)$-invariant 4-form $\sigma$ can be written as
\begin{equation*}
\sigma = y_1 d (\Sigma_1 \alpha_2 \alpha_3 )
+ y_2 d (\Sigma_2 \alpha_3 \alpha_1 )
+ y_3 d (\Sigma_3 \alpha_1 \alpha_2 ).
\end{equation*}

Finally, we want to show that $\sigma$ can be written
as $\omega^2 / 2$ for some two-form $\omega$, and that
$\omega \wedge \rho =0$. Explicitly, we can write
\begin{equation*}
\omega = \sqrt{{y_2 y_3 \over y_1}} \Sigma_1 \alpha_1
+ \sqrt{{y_1 y_3 \over y_2}} \Sigma_2 \alpha_2
+ \sqrt{{y_1 y_2 \over y_3}} \Sigma_3 \alpha_3.
\end{equation*}

It is straightfoward to check that this form $\omega$
satisfies the required properties, namely
\begin{equation*}
\sigma = {1 \over 2} \omega^2,
\end{equation*}
and
\begin{equation*}
\omega \wedge \rho =0.
\end{equation*}

The last thing we need to check before we proceed to the Hamiltonian
flow is to make sure that $x_i (t)$ and $y_i (t)$ are conjugate
coordinate and momenta. In other words, we need to show that
there is a non-degenerate pairing between invariant 3-forms and 4-forms:
$$
\langle \Sigma_2 \alpha_2 \Sigma_3 \alpha_3, d ( \Sigma_1 \alpha_1) \rangle
= \int_{{S}^3 \times T^3}
\Sigma_2 \alpha_2 \Sigma_3 \alpha_3 \Sigma_1 \alpha_1
= {\rm vol} ({S}^3) ~{\rm vol} (T^3) \ne 0.
$$
Therefore, just as in the model with $SU(2) \times SU(2)$
principal orbits \cite{Hitchin,BGGG},
it turns out that the symplectic form is a multiple of
$$
dx_1 \wedge dy_1 + dx_2 \wedge dy_2 + dx_3 \wedge dy_3.
$$

Using (\ref{defvor}) we find a simple expression for $V(\rho)$:
$$
V^2 (\rho) = - m^2 n^2 - 4m x_1 x_2 x_3.
$$
Since both $V (\rho)$ and $V (\sigma)$ must be real,
we have two constraints:
\begin{equation}
y_1 y_2 y_3 > 0, \quad 4m x_1 x_2 x_3 < - m^2 n^2.
\end{equation}
Provided these relations are satisfied, we can write
the Hamiltonian
\begin{eqnarray}
H & = & V(\rho) - 2 V(\sigma) \nn \\
& = & \sqrt{- m^2 n^2 - 4m x_1 x_2 x_3} - 2 \sqrt{y_1 y_2 y_3},
\label{hamilton}
\end{eqnarray}
which is constrained by the hypothesis of
Theorem \ref{hitthm} to be zero.

Now choosing $i\neq j\neq k\neq i$ among
${1,2,3},$
the corresponding Hamiltonian flow equations read
\begin{eqnarray}
\dot y_i & = & {2m x_j x_k \over \sqrt{- m^2 n^2 - 4m x_i x_j x_k}}, \nn \\
\dot x_i & = & - \sqrt{{y_j y_k \over y_i}}.
\label{odesystem}
\end{eqnarray}

A solution to these first-order differential equations
define a $G_2$ structure on $(a,b) \times {S}^3 \times T^3$.
Explicitly, the associative 3-form is given by
\begin{eqnarray}
\Phi = dt \wedge \left(
\sqrt{{y_2 y_3 \over y_1}} \Sigma_1 \alpha_1
+ \sqrt{{y_1 y_3 \over y_2}} \Sigma_2 \alpha_2
+ \sqrt{{y_1 y_2 \over y_3}} \Sigma_3 \alpha_3 \right) + \nn \\
+ n \Sigma_1 \Sigma_2 \Sigma_3 - m \alpha_1 \alpha_2 \alpha_3
+ x_1 d ( \Sigma_1 \alpha_1) + x_2 d ( \Sigma_2 \alpha_2)
+ x_3 d ( \Sigma_3 \alpha_3).
\label{calibr}
\end{eqnarray}
It follows that for $n=1$ the 3-sphere $B=SU(2)$ is
an associative submanifold inside $X=(a,b) \times SU(2) \times T^3$,
while the non-compact fiber $(a,b) \times T^3$ is coassociative.

Moreover, from the expression (\ref{calibr}) for the associative
3-form $\Phi$, it follows that the volume of $S^3$ and $T^3$
(measured with respect to the $G_2$-holonomy metric (\ref{hitmet})
obtained from $\Phi$) is bounded below:
\begin{eqnarray}
{\rm Vol} (S^3)& \geq &\vert n \vert \label{nbound},\\
{\rm Vol} (T^3)& \geq &\vert m \vert \label{mbound}.
\end{eqnarray}

In general, the $G_2$-holonomy metric looks like
\begin{eqnarray}
ds^2 = {\rm det} (K_{\Phi})^{-1/9}
\Big( \sqrt{y_1 y_2 y_3} dt^2
+ \sqrt{ {y_2 y_3 \over y_1} }
(x_2 x_3 \Sigma_1^2 + mn \Sigma_1 \alpha_1 - m x_1 \alpha_1^2) + \nn \\
+ \sqrt{ {y_1 y_3 \over y_2} }
(x_1 x_3 \Sigma_2^2 + mn \Sigma_2 \alpha_2 - m x_2 \alpha_2^2)
+ \sqrt{ {y_1 y_2 \over y_3} }
(x_1 x_2 \Sigma_3^2 + mn \Sigma_3 \alpha_3 - m x_3 \alpha_3^2) \Big)
\label{gmetric}
\end{eqnarray}
where
\begin{equation}
{\rm det} (K_{\Phi})
= - m^3 (y_1 y_2 y_3)^{3/2}
\Big(x_1 x_2 x_3 + {1 \over 4} m n^2 \Big)^3
= (y_1 y_2 y_3)^{9/2},
\end{equation}
where in the last equality we used the conservation of
the Hamiltonian, $H=0$.
Then, the overall factor ${\rm det} (K_{\Phi})^{-1/9}$
in the metric (\ref{gmetric}) cancels the coefficient
in front of $dt^2$, so that the resulting expression looks like
\begin{eqnarray}
ds^2 = dt^2
+ {1 \over y_1}
(x_2 x_3 \Sigma_1^2 + mn \Sigma_1 \alpha_1 - m x_1 \alpha_1^2) + \nn \\
+ {1 \over y_2}
(x_1 x_3 \Sigma_2^2 + mn \Sigma_2 \alpha_2 - m x_2 \alpha_2^2)
+ {1 \over y_3}
(x_1 x_2 \Sigma_3^2 + mn \Sigma_3 \alpha_3 - m x_3 \alpha_3^2).
\label{ggmetric}
\end{eqnarray}

\subsection{$SU(2)$ Symmetric Solution and Large Distance Asymptotics}

Let us study various limits of the new $G_2$ metric,
and try to understand the role of various parameters,
$m$ and $n$, in particular.
It is instructive to look first at the simple case, where:
\begin{equation}
x_1 = x_2 = x_3, \quad y_1 = y_2 = y_3.
\label{threexys}
\end{equation}
This set of extra conditions restricts us to a class of
metrics with extra $SU(2)$ symmetry.
As will be shown below, a study of much simplier $SU(2)$-invariant
metrics illustrates all important properties of
the generic solutions to (\ref{odesystem}).

The extra conditions (\ref{threexys}) lead to a consistent
truncation of the first-order system (\ref{odesystem}),
{\it cf.} \cite{gary} and \cite{BGGG}:
\begin{eqnarray}
\dot x &=& - \sqrt{y}, \nn \\
\dot y &=& {2mx^2 \over \sqrt{- m^2 n^2 - 4m x^3}}.
\label{sutwosys}
\end{eqnarray}
Without loss of generality, we can assume that $m$ is positive.
It implies that the values of $x(t)$ and $y(t)$ range in
\begin{eqnarray}
- \infty < & x(t) & \le - \left( {mn^2 \over 4} \right)^{1/3}, \nn \\
0 \le & y(t) & < + \infty.
\label{xyrange}
\end{eqnarray}

Since the system (\ref{sutwosys}) is Hamiltonian,
we have one obvious integral of motion,
namely $H (t)=0$ which we express in the form
\begin{equation}
- m x^3 = y^3 + {1 \over 4} m^2 n^2.
\label{hconst}
\end{equation}

This allows to reduce the system (\ref{sutwosys}) to a single
differential equation of a hypergeometric type,
\begin{equation}
{dy \over dt} = {m \over y^{3/2}}
\Big({1 \over m} y^3 + {1 \over 4} mn^2 \Big)^{2/3},
\label{singleode}
\end{equation}
which is exactly solvable:
\begin{equation}
t = t_0 + {1 \over 5} \Big( {2^7 \over m^5 n^4} \Big)^{1/3} y^{5/2}
F\left( \Big[ {5 \over 6} , {2 \over 3} \Big] , \Big[ {11 \over 6} \Big],
- {4 y^3 \over m^2 n^2}\right).
\end{equation}
This solution leads to a simple $G_2$-holonomy metric:
$$
ds^2 =  dt^2 + {1 \over y} \sum_{j=1}^3
(x^2 \Sigma_j^2 + mn \Sigma_j \alpha_j - m x \alpha_j^2),
$$
with the isometry group:
$$
SU(2) \times SU(2) \times U(1)^3.
$$

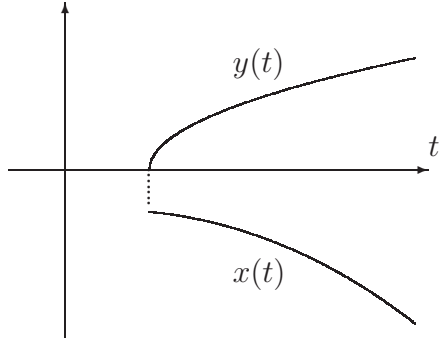
\begin{figure}
\setlength{\unitlength}{0.9em}
\begin{center}
\begin{picture}(22,11)

\put(5,5){\vector(1,0){15}}\put(20,5.5){$t$}
\put(7,-1){\vector(0,1){12}}
\qbezier(10,5)(10,7)(19.5,9)\put(13,8.5){$y(t)$}
\qbezier(10,3.5)(15,3)(19.5,-0.5)\put(13,1){$x(t)$}
\multiput(10,5)(0,-0.2){7}{\circle*{.1}}

\end{picture}\end{center}
\caption{The behavior of functions $x(t)$ and $y(t)$.}
\label{xyfig}
\end{figure}

The behavior of $x(t)$ and $y(t)$ is sketched in Fig. \ref{xyfig}.
Next, let us analyze various limits of this metric.

\subsubsection*{Solution with $n=0$ and Large Distance Asymptotics}

In the special case $n=0$, the solution takes a very simple
polynomial form:
\begin{equation}
x = - {m^{1/3} \over 4} (t - t_0)^2,
\quad y = {m^{2/3} \over 4} (t - t_0)^2.
\label{asympt}
\end{equation}

The corresponding metric looks like
(for simplicity, we put the integration constant $t_0=0$)
\begin{equation}
ds^2 = dt^2 + {1 \over 4} t^2 (\Sigma_1^2 + \Sigma_2^2 + \Sigma_3^2)
+ m^{2/3} (\alpha_1^2 + \alpha_2^2 + \alpha_3^2).
\label{trmetric}
\end{equation}
A numerical factor $1/4$ in front of the $\Sigma_i$ terms
is crucial here for these terms to become the usual Einstein
metric on the round 3-sphere. Hence, the above expression is
nothing but the usual Ricci-flat metric on
\begin{equation}
\mathbb{R}^4 \times {T}^3.
\label{trfour}
\end{equation}

The metric on the regular 3-torus in this solution depends
on the value of the parameter $m$. Namely, an easy computation
gives the asymtotic volume of the ${T}^3$ in this metric:
\begin{equation}
{\rm Vol} ({T}^3) \vert_{t=\infty} = m,
\end{equation}
which is in complete agreement with the bound (\ref{mbound}).
It is natural to expect that more general solutions, without
the $SU(2)$ symmetry, exhibit a similar behavior.
In the next subsection we will show that this is indeed the case;
changing various parameters it is easy to modify the asymptotic shape
of the 3-torus, but not the overall volume, which is determined only by $m$.
The metric (\ref{trmetric}) is manifestly Ricci-flat for all values
of the parameter $m$, so it is clearly a modulus.

Another remark is that (\ref{trmetric}) describes the asymptotic
behavior of the metric with non-zero $n$ at large distances.
Indeed, if the absolute value of $y(t)$
(and, therefore, of $x(t)$ as well) grows as $t \to \infty$,
the term with $n^2$ in the first order equation (\ref{singleode})
can be neglected and one finds a simple ODE:
$$
{dy \over dt} = m^{1/3} \sqrt{y},
$$
which leads to the {\it approximate} solution (\ref{asympt}).
Therefore, even for $SU(2)$ symmetric solutions with
non-zero parameter $n$, the metric is asymptotic to
$\mathbb{R}^4 \times {T}^3$ at large distances.
Of course, for $n \ne 0$ the metric can no longer have global
topology of $\mathbb{R}^4 \times {T}^3$ because the volume
of the 3-sphere is bounded by $\vert n \vert$, {\it cf.} (\ref{nbound}).
Nevertheless, as we pointed out, the parameter $n$
does not change the asymptotic behaviour of the metric.


\subsection{$U(1)$ Symmetric Solution and Short Distance Asymptotics}

We have studied the
asymptotic behavior of the $SU(2)$ symmetric solution
when the functions $x(t)$ and $y(t)$ approach one limit in
the range of allowed values (\ref{xyrange}). This limit corresponds
to large distance asymptotics of the $G_2$-holonomy metric.
Here, we discuss the other limit,
$$
x \to - \Big( {m n^2 \over 4} \Big)^{1/3}, \quad y \to 0,
$$
as $t$ approaches some value, say $t \to 0$.

In the special case $n=0$, we have also seen that the metric
is non-singular in this limit since the principal orbit
$M=S^3 \times T^3$ degenerates into $T^3$ at $t=0$
in such a way that the total space has topology (\ref{trfour}):
$$
X \cong \mathbb{R}^4 \times T^3
$$
On the other hand, if $m \ne 0$ and $n \ne 0$
the constraints (\ref{nbound}) and (\ref{mbound})
prevent both $S^3$ and $T^3$ cycles from shrinking.
In such cases one finds a rather exotic metric of the form:
$$
ds^2 = dt^2 + t^{- 2/5} \sum_{j=1}^3 ( \Sigma_j + \alpha_j )^2 + \ldots
$$
where the dots stand for the terms vanishing in the limit $t \to 0$.

%

In order to find other $G_2$-holonomy metrics with more regular
behavior at $t=0$, we have to allow some cycle to collapse
and relax the $SU(2)$ symmetry condition.
The natural step to consider between $SU(2)$ symmetry and
no symmetry at all is when only a
$U(1) \subset SU(2)$ symmetry
group is preserved. This can be achieved, for example,
by imposing the following conditions:
$$
x_1 = x_2, \quad y_1 = y_2.
$$
The corresponding expression for the metric (\ref{ggmetric}) looks like:
\begin{eqnarray}
ds^2 &=& dt^2
+ {x_1 x_3 \over y_1} (\Sigma_1^2 + \Sigma_2^2)
+ {mn \over y_1} (\Sigma_1 \alpha_1 - \Sigma_2 \alpha_2)
- {m x_1 \over y_1} ( \alpha_1^2 + \alpha_2^2) + \nn \\
&+& {1 \over y_3} (x_1^2 \Sigma_3^2 + mn \Sigma_3 \alpha_3 - m x_3 \alpha_3^2).
\label{uonemetric}
\end{eqnarray}

This leads to a rather simple system of only four
first-order equations:
\begin{eqnarray}
\dot x_1 = - \sqrt{y_3}, &&
\dot y_1 = {m x_1 x_3 \over y_1 \sqrt{y_3}}, \nn \\
\dot x_3 = - {y_1 \over \sqrt{y_3}}, &&
\dot y_3 = {m x_1^2 \over y_1 \sqrt{y_3}}.
\label{uonesystem}
\end{eqnarray}

In some special cases, this system has simple explicit solutions.
For example, if $n=0$ one finds
\begin{equation}
t = t_0 - x_1 \Big(- {B^2 \over \beta m}\Big)^{1/4}
F\left( \Big[ {1 \over 2} , {1 \over 4} \Big] , \Big[ {3 \over 2} \Big],
- B x_1^2 / \beta \right),
\end{equation}
where
\begin{equation}
\beta = x_3 y_3 - x_1 y_1
\label{uonebeta}
\end{equation}
and $B \le 0$ is assumed to be non-zero.
In the case $B=0$ we recover the $SU(2)$ symmetric solution (\ref{asympt}).
Notice, $\beta$ is an integral of motion. In the next subsection we
explain this in more detail and find all the integrals of motion
for the general first-order system (\ref{odesystem}).

However, trying to find a general solution to the reduced
first-order system (\ref{uonesystem}) amounts,
essentially, to solving the general system (\ref{odesystem}).
It can be written in terms of the Weierstrass function
and will be studied in the next subsection.
Here, let us consider approximate solutions at $t \to 0$.
There are many possibilities corresponding to different
assumptions about the
vanishing of the functions $x_i$ and $y_i$ at $t=0$.
We consider just one such possibility corresponding to
a solution, where the
$S^3$ cycle degenerates into a two-sphere:
$$
{S}^3 \to {S}^2, \quad t \to 0.
$$
Of course, we also have to assume $n=0$ in order to obey
the condition (\ref{nbound}).
Specifically, this solution corresponds to
$x_1 (t)$ and $y_1 (t)$ vanishing at $t=0$,
whereas $x_3 (t)$ and $y_3 (t)$
are assumed to take finite non-zero values at this point.
Therefore, from the integral of motion (\ref{uonebeta})
we get ($\beta < 0$)
\begin{equation}
y_3 \approx {\beta \over x_3},
\label{uoneapprox}
\end{equation}
and the conservation of the Hamiltonian (\ref{hamilton}) gives
$$
y_1 \approx \sqrt{ - {m \over \beta}} x_1 x_3
$$
For $x_1$ and $x_3$ we find a simple system of the first-order
differential equations, which has a nice trigonometric solution:
\begin{eqnarray}
x_1 = - A \sin (\gamma t), &&
y_1 =  {\sqrt{-m \beta} \sin (\gamma t) \over \gamma^2 A \cos^2 (\gamma t)},
\nn \\
x_3 = {\beta \over \gamma^2 A^2 \cos^2 (\gamma t)}, &&
y_3 = \gamma^2 A^2 \cos^2 (\gamma t),
\end{eqnarray}
with
$$
\gamma^2 = \sqrt{{m \over 4 |\beta|}}.
$$

The approximation (\ref{uoneapprox}) is valid
for $|\beta| \gg |x_1 y_1|$, which means
$$
\tan^2 (\gamma t) \ll 2
$$
In particular, this includes the interesting range of small $t$,
where the solution approximately behaves as
\begin{eqnarray}
x_1 \approx - A \gamma t, &&
y_1 \approx {\sqrt{-m \beta} t \over \gamma A },
\nn \\
x_3 \approx {\beta \over \gamma^2 A^2 }, &&
y_3 \approx \gamma^2 A^2.
\end{eqnarray}
The corresponding metric looks like
\begin{eqnarray}
ds^2 \approx dt^2
+ \sqrt{{|\beta| \over m}} (\Sigma_1^2 + \Sigma_2^2)
+ {m A^2 \over 2 |\beta|} ( \alpha_1^2 + \alpha_2^2)
+ t^2 \Sigma_3^2 + {4 \beta^2 \over A^4} \alpha_3^2.
\label{stwometric}
\end{eqnarray}
It is natural to expect that for $\gamma t \gg 1$
this metric interpolates to the asymptotic metric (\ref{asympt}).
Then, we obtain a smooth manifold $X$ with $G_2$ holonomy, such that
\begin{equation}
X \cong {S}^2 \times \mathbb{R}^2 \times T^3,
\label{stwosoln}
\end{equation}
where the volume of the two-sphere is proportional to $\sqrt{|\beta|}$.
In the next subsection we verfy that our expectation is correct
by constructing the explicit solution to the general first-order
system (\ref{odesystem}).


\subsection{General Solution}

Given
$$H = \sqrt{-m^2n^2 -4mx_1x_2x_3} - 2\sqrt{y_1y_2y_3}
= V (\rho) - 2 V (\sigma),$$
we recall that $H=0,$ i.e. $V(\rho) = 2 V(\sigma),$
along our orbit. Then consider the Hamiltonian
$$\widetilde{H} = V(\rho)^2 - 4 V(\sigma)^2
= -m^2n^2-4mx_1x_2x_3-4y_1y_2y_3.$$
Since $d(V(\rho)^2 - 4 V(\sigma)^2)
= 2V(\rho)(d V(\rho) - 2 d V(\sigma))$ along the
orbit, the orbits are the same, though parametrized differently.
Indeed,
\begin{equation}
\label{repar}
\frac{dt}{d\widetilde{t}} = 2 V(\rho) = 4 V(\sigma) = 4\sqrt{y_1y_2y_3}.
\end{equation}

Now the action $(\vec{x},\vec{y}\,)
\mapsto (M\cdot\vec{x},M^{-1}\cdot\vec{y}\,),$
where $M$ is a diagonal matrix with determinant one,
is symplectic and leaves the Hamiltonian invariant.
This symmetry group is two dimensional, and
the corresponding conserved quantities are
$$x_1y_1 - x_3y_3 \qquad \hbox{and} \qquad x_2y_2 - x_3y_3.$$

Now choose $i\neq j\neq k\neq i.$  The equations
of
motion are
$$\dot{x_i} = -4y_j y_k; \qquad
\dot{y_i} = 4mx_jx_k$$
where we now are using a dot over a variable
to represent $\frac{d}{d\widetilde{t}}.$
Let's define
$$z_i = x_i y_i.$$
Then $\frac{d}{d\widetilde{t}}
(z_i - z_j) = 0,$ as these are our
conserved quantities.  Therefore, we may write
\begin{eqnarray}
z & \equiv & z_1 \nn\\
z_2 & = & z + \alpha \label{abconst}\\
z_3 & = & z + \beta. \nn
\end{eqnarray}
Now define
$$X \equiv x_1 x_2 x_3 \qquad Y \equiv y_1 y_2 y_3.$$
Since our Hamiltonian is zero on our orbit, we have
\begin{equation}
\label{hiszero}
m^2 n^2 + 4m X + 4Y = 0.
\end{equation}

Now compute
$$\dot{z} = \dot{x_1}y_1 + x_1 \dot{y_1}=
-4Y + 4mX = -m^2 n^2 - 8Y.$$
But note
$$\dot{Y} = 4m(x_2x_3y_2y_3 + x_1x_3y_1y_3 + x_1x_2y_1y_2),$$
so we see
\begin{equation}
\label{zeq}
\ddot{z} = -32m(3z^2 + 2z(\alpha+\beta) + \alpha\beta).
\end{equation}

Assuming we can solve this equation, we can get
\begin{equation}
Y = -(\dot{z} + m^2n^2)/8,
\label{yyyviaz}
\end{equation}
and we can find the individual $x_i$ and $y_i$
as follows:
$\dot{x_1} = -4y_2y_3 = -4Y/y_1 = -4(Y/z)x_1,$
so we see
\begin{equation}
x_1 = - A_1\exp[-4\int (Y/z)\, d\widetilde{t}\,],
\label{xviaz}
\end{equation}
with $A_1$ a (positive) constant.
Then $y_1 = z/x_1.$  The quantities $x_2, y_2, x_3,
y_3$ are similarly calculated, with one of the integration
constants fixed by the constraint (\ref{hiszero}).
Finally, to connect with the form of the metric in (\ref{gmetric})
we must rewrite our answers in terms of $t,$
which means solving (\ref{repar}):
$$\frac{dt}{d\widetilde{t}} = 4\sqrt{Y}.$$

\subsubsection*{\bf Equation for $z$}

So let's try to solve the equation (\ref{zeq}).
First note that by
a change of variables
$$
u = -96m z - 32m(\alpha + \beta)
$$
the equation takes the form
\begin{equation}
\ddot{u} = u^2 + D
\label{ueq}
\end{equation}
where
$$D = (32m)^2(3\alpha\beta - (\alpha+\beta)^2)=
-(32m)^2(\alpha^2 - \alpha\beta + \beta^2).$$
We can try to solve this equation with a solution
satisfying a first-order equation
$\dot{u} = f(u).$
Then $\ddot{u} = f'\dot{u} = ff',$
so
$$(f^2)' = 2(u^2 + D),$$
and $f^2 = \frac{2}{3}u^3 + Du + E.$
Therefore,
$f = \pm\sqrt{\frac{2}{3}u^3 + Du + E}$
and we see
$\dot{u} = f(u)$ has the solution
$$\widetilde{t}
= C \pm\int\frac{du}{\sqrt{\frac{2}{3}u^3 + Du + E}},$$
which is the equation for the integral of an abelian
differential on the elliptic curve
$$y^2 = \frac{2}{3}x^3 + Dx + E.$$
Therefore, to express $u$ in terms of $\widetilde{t}$
we need to invert this (Abel-Jacobi) map.
This is precisely what the Weierstrass function does!

To make this explicit, we'll want to put things in
Weierstrass form.  So let's make the change of variables
$v = 6u$ and define
$$g_2 \equiv -D/3.$$
The equation for $v$ becomes
\begin{equation}
\ddot{v} = 6v^2 - g_2/2
\label{veq}
\end{equation}
with the solutions
$$\widetilde{t}
= C \pm \frac{dv}{\sqrt{4v^3 -
g_2v - g_3}}$$
where $C$ and $g_3$ are constants.
To invert this, we simply use the Weierstrass function
$$v = {\mathfrak p}_\tau (\widetilde{t}+C),$$
where $\tau$ is the modular parameter determined
by $g_2$ and the constant $g_3.$  Note that this
is a two-parameter family of solutions
(since the Weierstrass function is even, we don't
gain anything by including the solution with
$-\widetilde{t}+C$
as the argument).
To be sure, note that $\mathfrak{p}$
famously satisfies the differential equation
$$(\mathfrak{p'})^2 = 4\mathfrak{p}^3 - g_2 \mathfrak{p} - g_3.$$
Differentiating once more, we see
$$2\mathfrak{p}'\mathfrak{p}'' = 12\mathfrak{p}^2\mathfrak{p}'
- g_2\mathfrak{p}',$$
so
$$\mathfrak{p}'' = 6\mathfrak{p}^2 - g_2/2.$$

Since the lattice of the torus is rectangular when
$g_2$ and $g_3$ are real, the Weierstrass function
has complex conjugation symmetry, i.e. it's real.

For some special examples, when $g_2 = 8$,
we have the simple solution
$$v = \csc^2(\widetilde{t}+C),$$
and if $g_2 = 4/3$ we have the
solution $$v = \csc^2(\widetilde{t}+C) - 1/3.$$

\subsubsection*{Asymptotics and Behavior Near the Poles}

Note that the Weierstrass $\mathfrak{p}$-function has
a second-order pole. Below we argue that it corresponds
to the asymptotic region of our solution, where the metric
has the same asymptotic $\mathbb{R}^4 \times {T}^3$ behavior
as in (\ref{trfour}).

Specifically, near a pole we have
$$
v \approx \widetilde{t}^{-2}.
$$
In order to verify that this gives an approximate
solution to the eqn. (\ref{veq}), note that for large $v$
the constant term $g_2/6$ can be neglected.
Unwinding the definitions gives (omitting the subleading terms)
$$
z \approx - {u \over 96 m} \approx - {v \over 576 m}
\approx - {1 \over 576 m \widetilde{t}^2}.
$$
Therefore, $x_i (\widetilde{t})$ and $y_i (\widetilde{t})$ look like
$$
x_i \sim y_i \sim {1 \over \widetilde{t}},
$$
and
$$
Y \approx {1 \over 2304 m \widetilde{t}^3}.
$$

Using the relation (\ref{repar}) between $t$ and $\widetilde{t}$,
we find the asymptotic solution in terms of the original variable $t$:
\begin{eqnarray}
& x_1 = - A_1 t^2, & y_1 = {m \over 16 A_1} t^2, \nn\\
& x_2 = - A_2 t^2, & y_2 = {m \over 16 A_2} t^2, \label{generalxy} \\
& x_3 = - A_3 t^2, & y_3 = {m \over 16 A_3} t^2,
\end{eqnarray}
where we assume that the integration constants $A_i$
are positive (to agree with our earlier conventions $x_i < 0$).
As we pointed out earlier, one of the integration constants
is not independent due to the constraint (\ref{hiszero}), which reads
\begin{equation}
64 A_1 A_2 A_3 = m.
\label{hzeroviaa}
\end{equation}
Hence, we can eliminate one of the constants $A_i$, say, $A_3$:
$$
A_3 = {m \over 64 A_1 A_2}.
$$

It is easy to check directly that (\ref{generalxy}) is an asymptotic
solution to the original first order system (\ref{odesystem}).
In the limit when $x_i$ are large (equivalently, in the limit $n \to 0$),
the equations look like
\begin{equation}
\dot y_i = \sqrt{ - {m x_j x_k \over x_i}}, \quad
\dot x_i = - \sqrt{{y_j y_k \over y_i}}.
\end{equation}
Analysis, similar to what we have done in the $SU(2)$-invariant case,
shows that $x_i$ and $y_i$ have
to grow as $t^2$, and straightfoward
calculation leads to the solution (\ref{generalxy}).

Now, substituting (\ref{generalxy}) into (\ref{ggmetric}),
we find the (asymptotic) expression for
the $G_2$-holonomy metric:
\begin{equation}
ds^2 = dt^2 +
{1 \over 4} t^2 (\Sigma_1^2 + \Sigma_2^2 + \Sigma_3^2)
+ 16 (A_1^2 \alpha_1^2 + A_2^2 \alpha_2^2 + A_3^2 \alpha_3^2).
\label{nzerometr}
\end{equation}
Note, that the first terms describe the usual metic on $\mathbb{R}^4$.
The coefficient $1/4$ is crucial for this and comes as follows:
$$
{x_1 x_2 \over t^2 y_3} = A_1 A_2 \Big( {m \over 16 A_3} \Big)^{-1}
= {16 A_1 A_2 A_3 \over m} = {1 \over 4}.
$$
Here we used the identity (\ref{hzeroviaa}) satisfied by $A$'s.

Therefore, the metric (\ref{nzerometr}) describes the flat metric on
$$
\mathbb{R}^4 \times T^3.
$$
However, unlike the $SU(2)$-symmetric metric (\ref{trmetric}),
the solution here describes a 3-torus of arbitrary shape
(determined by $A_1$ and $A_2$) with a fixed volume:
\begin{equation}
{\rm Vol} (T^3) = \int_{T^3} \sqrt{g} =
16^{3/2} A_1 A_2 A_3 = m,
\end{equation}
where we again used the important condition (\ref{hzeroviaa}).

To summarize, we have demonstrated that the general solution
is asymptotic to $\mathbb{R}^4 \times T^3$ where the Weierstrass
function has a second-order pole. Moreover, the size of the 3-torus
asymptotically saturates the bound (\ref{mbound}).

At this point we shall remark on the interpretation of various
parameters in the general solution described here.
In total we have eight parameters:  six integration constants
for the first-order system (\ref{odesystem}) and the original
parameters $m$ and $n$. One of the integration constants
(corresponding to the conservation of the Hamiltonian)
is a constant in the definition of the radial variable $t$,
and therefore does not play an interesting role.
The remaining five independent constants have been denoted
$\alpha$, $\beta$, $E$, $A_1$ and $A_2$.
Two of them, $A_1$ and $A_2$, affect the behavior of the metric
at infinity.
Namely, they describe the asymptotic form of the $T^3$.
Furthermore, the volume of the torus is determined by $m$.
On the other hand, the parameter $n$ along with the integration
constants $\alpha$, $\beta$, and $E$ should be interpreted
as `dynamical moduli' since they don't change behavior of
the metric at infinity. In particular, the value of $n$
determines the minimal volume of the 3-sphere ({\it cf.} (\ref{nbound})),
and if $n=0$ the value of $\beta$ determines the volume
of the two-sphere in (\ref{stwosoln}).


\section{Abelian BPS Monopoles from Torus Fibrations}
\label{monopole}

The analysis in this section is motivated by the
cosmic string solution \cite{GSVY}, where compactification
of Type-II string theory on a two-torus with varying
modulus $\tau$ was considered.
According to the equations of motion of the effective theory,
if $\tau = \tau (z)$ depends only on two real directions of
space-time (which can be combined in one
complex coordiante $z$),
it must be holomorphic.
Then, at the
points of space-time where $\tau \rightarrow \infty,$
a real codimension
two singularity --- a cosmic string --- is found.
In this way, one views fibrations by special-holonomy fibers
as supersymmetric topological defects in lower dimension.

Our solutions, constructed in the previous sections, do not have
degenerate fibers. Therefore, one would not expect to find
extreme concentration of energy via Kaluza-Klein reduction.
Nevertheless, we will explain below that
the dimensionally-reduced field configurations carry
topological charge.
So, they indeed represent stable solitons --- monopoles,
cosmic strings, or domain walls, depending on the configuration
of energy density.
The soliton obtained by Kaluza-Klein reduction
is guaranteed to be BPS because the original metric
admits a covariantly constant spinor.

Disk instanton corrections to the geometry near the cosmic string
singularities should smoothen the metric, and the authors
of \cite{GSVY} argued that a smooth, even compact, total
space could result.  This line of thinking was given
credence by the explicit metrics near a degeneration
found in Ooguri and Vafa in \cite{OV}.  While our metrics
involve the smooth part of a fibration,
we hope that similar effects involving associative
three-cycles will result in compact manifolds of
$G_2$ holonomy, fibered over an $S^3$ base which includes
the discriminant locus of a torus or $K3$ fibration
({\it cf.} \cite{GW}).

%
Since the base of a coassociative fibration is three-dimensional,
the reduced solution could be interpreted as a monopole,
after we supplement the metric on $X$ with time direction $\tau$:
$$
\mathbb{R}_{\tau} \times X.
$$
Of course, this space-time has the same holonomy as $X$,
so the solution is guaranteed to be supersymmetric.
Mainly interested in abelian monopoles, we shall focus
on the torus fibrations\footnote{Kaluza-Klein reduction of
certain $G2$ holonomy metrics to non-abelian
monopoles has been discussed recently in \cite{Hartnoll}.}
found in Sec. \ref{torusfib}.

The solutions we found can be classified according to their isometry.
After Kaluza-Klein reduction to $3+1$ dimension this translates to
the rotational symmetry of the monopole solution \cite{Hartnoll}.
Namely, the
generic metric (\ref{ggmetric}) is expected to give
a $(3+1)$-dimensional monopole metric with no rotational symmetry.
On the other hand, solutions with extra $SU(2)$ (resp. $U(1)$)
symmetry lead to spherically (resp. axially) symmetric monopoles.
We summarize this general pattern in Table 1.

\begin{table}\begin{center}
\begin{tabular}{|c|c|}
\hline
$G_2$ Manifold $X$ & Monopole Solution \\
\hline
\hline
$SU(2)$ Symmetry & Spherical Symmetry  \\
\cline{1-2}
$U(1)$ Symmetry & Axial Symmetry  \\
\cline{1-2}
No Extra Symmetry & No Rotational Symmetry \\
\hline
\end{tabular}\end{center}
\caption{Relation between extra symmetry of the $G_2$-holonomy
metric (\ref{ggmetric}) and rotational symmetry of the corresponding
monopole solution.}
\end{table}

In order to avoid possible confusion with a space-like coordinate $t$,
we introduce a time-like variable $\tau$ and replace $t$ by $r$,
to emphasize that it plays a role of radial variable.
Now, let us rewrite the eight-dimensional metric
on $\mathbb{R}_{\tau} \times X$ in the new notation:
\begin{eqnarray}
ds^2 = - d \tau^2 + dr^2
+ {1 \over y_1}
(x_2 x_3 \Sigma_1^2 + mn \Sigma_1 \alpha_1 - m x_1 \alpha_1^2) + \nn \\
+ {1 \over y_2}
(x_1 x_3 \Sigma_2^2 + mn \Sigma_2 \alpha_2 - m x_2 \alpha_2^2)
+ {1 \over y_3}
(x_1 x_2 \Sigma_3^2 + mn \Sigma_3 \alpha_3 - m x_3 \alpha_3^2),
\label{ggmmetric}
\end{eqnarray}
where $x_i$ and $y_i$ should be understood as functions
of the radial variable $r$:
\begin{eqnarray}
x_i &\equiv& x_i (r), \nn\\
y_i &\equiv& y_i (r).
\end{eqnarray}

Since the $G_2$-holonomy manifold $X$ has principal orbits
$SU(2) \times T^3$, it has four natural $U(1)$ isometries:
three from the isometries of the 3-torus, and a $U(1) \subset SU(2)$.
The latter is generated by shifts of the angular variable
$\phi$, {\it cf.} (\ref{sutwoforms}). In order to treat this
latter $U(1)$ in the same way as the three directions of the $T^3$,
it is convenient to introduce
$$
\alpha_4 \equiv d \phi.
$$
Then, $\alpha_i$, $i=1,2,3,4$ is a natural basis of one-forms
on the 4-torus, $T^4 = T^3 \times U(1)$.

Now, we are ready to make a Kaluza-Klein reduction on the $T^4$.
We write the metric in the usual Scherk-Schwarz form \cite{SS}
\begin{equation}
ds^2 = ds_{1,3}^2 + h_{ij} (\alpha_i + A_i) (\alpha_j + A_j),
\label{ssmetric}
\end{equation}
where $ds^2_{1,3}$ is the four-dimensional metric of the static
gravitating monopole solution
and $A_i$ is the gauge connection
for the $i$-th $U(1)$ gauge factor.
The dilaton-like scalar fields $h_{ij}$ have charge $+1$
under $i$-th $U(1)$ gauge factor, and $-1$ under the $j$-th $U(1)$.
All these fields appear in the appropriate supermultiplets
of the effective four-dimensional theory.
Summarizing, after the Kaluza-Klein reduction we find
the following spectrum of the effective supersymmetric theory
in four dimensions:

\begin{center}
{\bf 4D Theory:} Supergravity
coupled to 4 vector and 10 matter multiplets.
\end{center}

Straightforward but technical calculations give the scalar field
matrix corresponding to the general solution (\ref{ggmetric}):
\begin{equation}
h = \begin{pmatrix}
- {m x_1 \over y_1} & 0 & 0 & {mn \over 2 y_1} \sin \psi \sin \theta \cr
0 & - {m x_2 \over y_2} & 0 &{mn \over 2 y_2} \cos \psi \sin \theta \cr
0&0& - {m x_3 \over y_3} & {mn \over 2 y_3} \cos \theta \cr
{mn \over 2 y_1} \sin \psi \sin \theta &
{mn \over 2 y_2} \cos \psi \sin \theta &
{mn \over 2 y_3} \cos \theta & h_{44}
\end{pmatrix},
\label{hhmatrix}
\end{equation}
where
$$
h_{44} = {x_1 x_2 \over y_3} \cos^2 \theta + {x_3 \over y_1 y_2}
(x_1 y_1 \cos^2 \psi + x_2 y_2 \cos^2 \psi) \sin^2 \theta.
$$

For a general solution,
the gauge connections can be conveniently written as:
\begin{equation}
A_i = \sum_{k=1}^4 h^{-1}_{ik} \widetilde{A}_k,
\label{afields}
\end{equation}
where
\begin{eqnarray}
\widetilde{A}_1 & = & {mn \over 2 y_1} \cos \psi d \theta, \cr
\widetilde{A}_2 & = & - {mn \over 2 y_2} \sin \psi d \theta, \cr
\widetilde{A}_3 & = & {mn \over 2 y_3} d \psi, \cr
\widetilde{A}_4 & = & {x_1 x_2 \over y_3} \cos \theta d \psi
+ \Big( {x_2 x_3 \over y_1} - {x_1 x_3 \over y_2} \Big)
\cos \psi \sin \psi \sin \theta d \theta.
\label{tildafields}
\end{eqnarray}

Finally, the metric in $(3+1)$ dimensions looks like:
\begin{equation}
ds^2_{1,3} = - d \tau^2 + dr^2 + {x_1 x_2 \over y_3} d \psi^2
+ \Big( {x_2 x_3 \over y_1} \cos^2 \psi
+ {x_1 x_3 \over y_2} \sin^2 \psi \Big) d \theta^2
- \sum_{i,j=1}^4 A_i h_{ij}^{-1} A_j.
\label{longmetric}
\end{equation}
Evaluating the last term leads to a rather complicated
form of the metric, which we write explicitly only in
a few simple examples below.


\subsection{Spherically Symmetric Monopoles}

The above formulas considerably simplify in the case
of the $SU(2)$ symmetric solution (\ref{trmetric}):
$$
x_1 = x_2 = x_3, \quad y_1 = y_2 = y_3.
$$
For example,
%
the gauge connections (\ref{afields}) can be written explicitly:
\begin{eqnarray}
A_1 & = & - \Big( {n \over 2 x} \Big)
(\cos \psi d \theta - \cos \psi \sin \psi \sin
\theta d \psi), \cr
A_2 & = & \Big( {n \over 2 x} \Big)
(\sin \psi d \theta +
\cos \psi \cos \theta \sin \theta d \psi), \cr
A_3 & = & - \Big( {n \over 2 x} \Big) \sin^2 \theta d \psi, \cr
A_4 & = & \cos \theta d \psi.
\label{gaugefields}
\end{eqnarray}
Note, that $A_1$, $A_2$, and $A_3$ are all proportional to $n$,
unlike $A_4$.

Also, as we alluded to earlier,
the $SU(2)$ symmetric solution automatically leads
to the spherically symmetric monopole metric
(in the Einstein frame):
\begin{equation}
ds^2_E = \Big( - {mx \over y^{1/2}} \Big) (- d \tau^2 + dr^2 )
+ y^{3/2} d \Omega_2^2.
\label{xxxextmet}
\end{equation}

%
In the simple case $n=0,$
the metric on the $G_2$-holonomy manifold $X$
becomes the usual metric on $\mathbb{R}^3 \times T^3$.
Specifically, the functions $x(r)$ and $y(r)$
take a simple polynomial form (\ref{asympt}):
\begin{equation}
x = - {m^{1/3} \over 4} r^2, \quad y = {m^{2/3} \over 4} r^2.
\label{xypolyr}
\end{equation}

After reduction to $3+1$ dimensions, the scalar field
matrix (\ref{hhmatrix}) turns out to be diagonal:
\begin{equation}
h = {\rm diag} (m^{2/3}, m^{2/3}, m^{2/3}, \varphi ),
\quad \varphi = {1 \over 4} r^2,
\label{kkh}
\end{equation}
and the resulting metric is spherically symmetric:
\begin{equation}
ds^2_E = {1 \over 2} mr (- d \tau^2 + dr^2)
+ {m \over 8} r^3 d \Omega_2^2.
\label{kkds}
\end{equation}

What is particularly nice about this solution
is that the gauge fields $A_1$, $A_2$, and $A_3$ vanish in
this background, so that we end up with a localized particle,
magnetically charged under a single $U(1)$.
Furthermore, $A_4$ resembles the gauge connection of
the Pollard-Gross-Perry-Sorkin magnetic monopole \cite{PGPS}:
\begin{equation}
A_4 = \cos \theta d \psi.
\label{kka}
\end{equation}
In order to realize (\ref{kkh})--(\ref{kka}) as a solution
in the Kaluza-Klein theory, it is convenient to combine
the $(3+1)$-dimensional metric, the gauge field $A_4$,
and the ``dilaton'' $\varphi$ into
a five-dimensional metric.
Then the equations for all of these fields follow from
the five-dimensional supergravity action with the usual bosonic piece
$$
S = - {1 \over 16 \pi \kappa_5} \int d^5 x \sqrt{-g_5} R_5.
$$

Unlike the usual Kaluza-Klein monopole \cite{PGPS}, however,
this solution represents a distribution of the magnetic charge
in the entire three-dimensional space.
It follows, for example, from (\ref{kkh}) that the dilaton
field $\varphi$ has a uniform source in the $\mathbb{R}^3$.
The reason is that the size of the circle, parametrized
by $\phi$ grows at large distances, {\it cf.} (\ref{xypolyr}).
In order to obtain a solution with localized source, one
has to start with a metric (\ref{ggmmetric}), where
the functions $x_i x_j / y_k$ are bounded at large $r$.
This would work, for example, if we had a Taub-NUT space
instead of $\mathbb{R}^4$ in our special solution.
It is easy to check, however, that TN$^4 \times T^3$
is not among the metrics of the form (\ref{ggmetric})
that we consider here.

Another important observation is that $(3+1)$-dimensional
metric (\ref{longmetric}) always has
the simple asymptotic
behavior (\ref{kkds}) of a Kaluza-Klein magnetic monopole.
Indeed, even for a general solution without $SU(2)$ symmetry,
the asymptotic form of the metric (\ref{nzerometr}) describes
a flat metric on $\mathbb{R}^4 \times T^3$, where the shape
of $T^3$ can be different now.  After reduction to $3+1$
dimensions this may change the asymptotic vevs of the scalar
fields, but not the $(3+1)$-dimensional metric (\ref{kkds}).

\subsection{Axially Symmetric Monopoles}

Axially symmetric BPS monopoles follow from the $U(1)$ symmetric
solution, just like spherically symmetric ones follow from solutions
with additional $SU(2)$ isometry. The simplest way to see this is
to put $x_1 = x_2$ and $y_1 = y_2$
in the general expressions for the scalar fields (\ref{hhmatrix}),
$U(1)$ gauge connections (\ref{tildafields}), and monopole
metric (\ref{longmetric}).
For example, 
$$
A_4 = {z \cos \theta \over z + \beta \sin^2 \theta} d \psi,
$$
where, following the notation of section \ref{torusfib},
we use $z=x_1 y_1$.
Notice, that $A_4$ does not depend on the angular variable $\psi$,
indicating the axial symmetry of the four-dimensional solution.
The same is true for all the other fields.
Thus, in the Einstein frame the metric reads
\begin{eqnarray}
ds^2_E &=& \sqrt{{\rm det} ~h} \Big[
- d \tau^2 + dr^2 - {y_1^2 y_3 \over mz}
\Big( d \theta^2 + {z \sin^2 \theta \over z + \beta \sin^2 \theta}
d \psi^2 \Big) \Big] = \nn\\
&=&
{m \sqrt{z^2 + \beta z \sin^2 \theta} \over y_1 \sqrt{y_3}}
\Big( - d \tau^2 + d r^2  - {y_1^2 y_3 \over mz} d \theta^2 \Big)
- {y_1 \sqrt{y_3} z \sin^2 \theta \over
\sqrt{z^2 + \beta z \sin^2 \theta}} d \psi^2.\qquad
\label{dipole}
\end{eqnarray}
As expected, the metric is manifestly axially symmetric.
At large distances, $|z| \gg |\beta|$, the dependence on $\theta$
drops out and the metric takes the asymptotic, spherically
symmetric form (\ref{kkds}). This is also expected from
the general no-hair theorem.

On the other hand, at small distances, the magnetic source
is extended in one of the directions, thus breaking $SO(3) \cong SU(2)$
rotational symmetry down to $U(1)$ axial symmetry.
Therefore, it is natural to interpret such metric as
a magnetic monopole, which also carries some dipole charge
given by $\beta$. Indeed, as $\beta \to 0$ the solution
reduces to a spherically symmetric monopole.
A further argument for this interpretation is that
at large distances the field of a dipole falls off
much faster than the field of a monopole, in agreement with
the asymptotic behavior of the axially symmetric solution (\ref{dipole}).
Depending on the internal structure of the dipole, one might say
that it consists of two point-like sources connected by a finite string.

Let us consider a specific solution (\ref{stwometric}) studied
in the previous section. As $z \to 0$, the corresponding
four-dimensional metric looks like
$$
ds^2_E \approx {m \over \sqrt{2}} \sin \theta
\Big( - d \tau^2 + dr^2 + {2 |\beta| \over m} d \theta^2
+ 2 \sqrt{|\beta| \over m} r^2 d \psi^2 \Big).
$$
It has two cosmological singularities:
one at $\theta=0$, and another at $\theta = \pi$.

\vskip0.2in
\centerline{\bf Acknowledgments}
\vskip0.1in
We would like to thank A. Strominger, R. Thomas, C. Vafa,
and E. Witten for instructive discussions.
This research was conducted during the period S.G.
served as a Clay Mathematics Institute Long-Term Prize Fellow.
The work of S.G. is also supported in part by grant RFBR No. 01-02-17488,
and the Russian President's grant No. 00-15-99296.
The work of S.T.Y. is supported in part by grants
DMS-0074329 and DMS-9803347.
The work of E.Z. is supported in part by
NSF grant DMS-0072504 and by an Alfred P. Sloan
Foundation Fellowship.


\begin{thebibliography}{99}

\bibitem{BA}
B.S.~Acharya, ``Exceptional Mirror Symmetry,''
in {\sl Winter School on Mirror Symmetry,
Vector Bundles and Lagrangian Submanifolds,}
C. Vafa and S.-T. Yau, eds., AMS and
International Press, Boston, 2001.

\bibitem{Bobby}
B.S.~Acharya, ``On Mirror Symmetry for Manifolds of Exceptional Holonomy,''
Nucl.Phys. {\bf B524} (1998) 269.

\bibitem{AchW} {B. Acharya and E. Witten, ``Chiral Fermions
from Manifolds of $G_2$ Holonomy,'' hep-th/0109152.}

\bibitem{AMV}
M.~Atiyah, J.~Maldacena and C.~Vafa,
``An M-theory flop as a large N duality,'' hep-th/0011256.

\bibitem{AtW}
M.~Atiyah and E.~Witten,
``M-theory dynamics on a manifold of $G_2$ holonomy,'' hep-th/0107177.

\bibitem{BSV} M.~Bershadsky, V.~Sadov, C.~Vafa, ``D-Branes and Topological
Field Theories,'' Nucl. Phys. {\bf B463} (1996) 420.

\bibitem{BGGG}
A.~Brandhuber, J.~Gomis, S.~S.~Gubser and S.~Gukov,
``Gauge theory at large N and new G(2) holonomy metrics,''
hep-th/0106034.

\bibitem{gary}
R.~Bryant and S.~Salamon,
"On the Construction of some
Complete Metrics with Exceptional Holonomy",
{\em Duke Math.} {\bf J. 58} (1989) 829.

G. W.~Gibbons, D. N.~Page, C. N.~Pope, ``Einstein
Metrics on $S^3$, $\mathbb{R}^3$ and $\mathbb{R}^4$ Bundles,''
{\em Commun.Math.Phys} {\bf 127} (1990) 529--553.

\bibitem{Swann}
R.~Cleyton and A.~Swann, ``Cohomogeneity-one $G_2$ Structures,''
math.dg/0111056.

\bibitem{GSVY} B. Greene, A. Shapere, C. Vafa, and S.-T.
Yau, ``Stringy Cosmic Strings and Noncompact
Calabi-Yau Manifolds,'' Nucl. Phys. {\bf B337} (1990) 1--36.

\bibitem{Gross} M.~Gross, ``Topological
Mirror Symmetry,'' Invent. Math. {\bf 144} (2001) 75--137;
and ``Special Lagrangian Fibrations I:  Topology,''
in {\sl Winter School
on Mirror Symmetry, Vector Bundles and
Lagrangian Submanifolds,} C. Vafa and S.-T. Yau, eds.,
AMS/International Press (2001) 65--93.

\bibitem{GW} M. Gross and P. M. H. Wilson,
``Large Complex Structure Limits of $K3$ Surfaces,''
math.DG/0008018.

\bibitem{Gukov} S.~Gukov, ``Solitons, Superpotentials and Calibrations,''
Nucl.Phys. {\bf B574} (2000) 169.

\bibitem{Hartnoll} S. Hartnoll, ``Axisymmetric non-abelian
BPS monopoles from $G2$ metrics,'' hep-th/0112235.

\bibitem{HS} J. Harvey and A. Strominger, ``The Heterotic
String is a Soliton,'' Nucl. Phys. {\bf B449} (1995)
535--552; erratum---ibid. {\bf B458} (1996) 456--73.

\bibitem{H} N. Hitchin, ``The Moduli Space of Special
Lagrangian Submanifolds,''
Ann. Scuola Norm. Sup. Pisa Cl. Sci. (4) {\bf 25}
(1997) 503--515; math.DG/9711002.

\bibitem{Hit}
N.~Hitchin, ``The geometry of three-forms in six and seven dimensions,''
math.DG/0010054.

\bibitem{Hitchin}
N.~Hitchin,
``Stable forms and special metrics,'' math.DG/0107101.

\bibitem{KV} S. Kachru and C. Vafa, ``Exact Results
for $N=2$ Compactification of Heterotic Strings,''
Nucl. Phys. {\bf B450} (1995) 69--89.

\bibitem{Kob} R. Kobayashi,
``Moduli of Einstein Metrics on a K3
Surface and Degeneration of Type I,''
in {\sl Advanced Studies in Pure Mathematics 18--II:
K\"ahler Metric and Moduli Spaces,}
T. Ochiai, ed., Academic Press (1990)
257--311.

\bibitem{LeeLeung}
J.-H. Lee and N. C. Leung,
``Geometric Structures on $G_2$ and $Spin(7)$-Manifolds,''
math.DG/0202045.

\bibitem{Mayr} P. Mayr, ``On Supersymmetry Breaking in
String Theory and its Realization in Brane
Worlds,'' Nucl. Phys. {\bf B593} (2001) 99--126; hep-th/0003198.

\bibitem{M} R. McLean, ``Deformations of Calibrated
Submanifolds,'' Comm. Anal. Geom. {\bf 6} (1998)
705--747.

\bibitem{OV} H. Ooguri and C. Vafa, ``Summing Up
D-Instantons,'' Phys. Rev. Lett. {\bf 77} (1996) 3296--3298.

\bibitem{OYZ} H. Ooguri, Y. Oz, and Z. Yin,
``D-Branes on Calabi-Yau Spaces and Their
Mirrors,'' Nucl. Phys. {\bf B477} (1996) 407--430.

\bibitem{PT} G.~Papadopoulos and P.K.~Townsend,
``Compactification of D=11 supergravity on spaces of
exceptional holonomy,'' {\em Phys.Lett.} {\bf B357} (1995) 300.


\bibitem{PGPS} D.~Pollard, {\em J. Phys.} {\bf A16} (1983) 565;
D.J.~Gross, M.J.~Perry, {\em Nucl. Phys.} {\bf B226} (1983) 29;
R.D.~Sorkin, {\em Phys. Rev. Lett.} {\bf 51} (1983) 87.

\bibitem{SS} J. Scherk, J. Schwarz, ``How to get masses from
extra dimensions,'' Nucl. Phys. {\bf B153} (1979) 61.

\bibitem{SV}
S.L. Shatashvili and C. Vafa, ``Superstrings and Manifolds of
Exceptional Holonomy,'' hep-th/9407025.

\bibitem{SYZ} A.~Strominger, S.-T.~Yau, and
E.~Zaslow, ``Mirror Symmetry is T-Duality,''
Nucl. Phys. {\bf B479} (1996) 243--259.

\bibitem{TV} T. Taylor and C. Vafa, ``RR Fulx
on Calabi-Yau and Partial Supersymmetry
Breaking,'' Phys. Lett. {\bf B474} (2000) 130--137.

\bibitem{T}
W. Thurston,
``Three-Dimensional Manifolds,
Kleinian Groups and Hyperbolic Geometry,''
Bull. Amer. Math. Soc. (N.S.) {\bf 6} (1982) 357--381.

\bibitem{VW} C.~Vafa, E.~Witten, ``A Strong Coupling Test
of S-Duality,'' Nucl. Phys. {\bf B431} (1994) 3.

\bibitem{W} E. Witten, ``Five-Brane Effective Action
in M-Theory,'' J. Geom. Phys. {\bf 22} (1997) 103--133.



\end{thebibliography}
\end{document}